# Nanopore Fabrication by Controlled Dielectric Breakdown

Harold Kwok*, Kyle Briggs*, and Vincent Tabard-Cossa

*Center for Interdisciplinary Nanophysics, Department of Physics, University of Ottawa, 150 Louis Pasteur, Ottawa, Ontario K1N 6N5, Canada*

** These authors contributed equally.*

*Corresponding author:* Vincent Tabard-Cossa, tcossa@uOttawa.ca

*Keywords:* Nanopore, Nanofabrication, Dielectric Breakdown, DNA translocation, Bio-nanotechnology

## ABSTRACT

Nanofabrication techniques for achieving dimensional control at the nanometer scale are generally equipment-intensive and time-consuming. The use of energetic beams of electrons or ions has placed the fabrication of nanopores in thin solid-state membranes within reach of some academic laboratories, yet these tools are not accessible to many researchers and are poorly suited for mass-production. Here we describe a fast and simple approach for fabricating a single nanopore down to 2-nm in size with sub-nm precision, directly in solution, by controlling dielectric breakdown at the nanoscale. The method relies on applying a voltage across an insulating membrane to generate a high electric field, while monitoring the induced leakage current. We show that nanopores fabricated by this method produce clear electrical signals from translocating DNA molecules. Considering the tremendous reduction in complexity and cost, we envision this fabrication strategy would not only benefit researchers from the physical and life sciences interested in gaining reliable access to solid-state nanopores, but may provide a path towards manufacturing of nanopore-based biotechnologies.

## SIGNIFICANCE STATEMENT

The main barrier to further development of nanopore-based technologies is the complexity, low-throughput, and high-cost associated with of current nanofabrication techniques relying on beams of high-energy particles. New fabrication strategies are needed for the field to continue to thrive and the promised health-related applications to be successfully commercialized. In this paper, we present an original fabrication method for creating nanopores directly in aqueous salt solutions, achieving a ~$10^6$ reduction in instrumentation cost and significantly higher yield with an automated fabrication process. We envision this fabrication strategy will not only provide a path towards nanomanufacturing of nanopore-based devices, but will democratize the use of solid-state nanopores, while offering researchers new strategies for integrating nanopores with CMOS and microfluidics technologies.

## INTRODUCTION

Nanopore sensing relies on the electrophoretically driven translocation of biomolecules through nanometer-scale holes embedded in thin insulating membranes to confine, detect and characterize the properties or the activity of individual biomolecules electrically, by monitoring transient changes in ionic current (1–4). The field was initially shaped by the ability of researchers to exploit biological channels to translocate single molecules (5, 6). It rapidly expanded when new techniques to fabricate individual molecular-sized holes in thin solid-state materials were developed over the last decade (7–12). These techniques, based on beams of high-energy particles, either produced by a dedicated ion beam machine (i.e ion-beam sculpting) or a transmission electron microscope (i.e TEM drilling), allowed researchers to control the nanopore size at the sub-10-nm length scale with single nanometer precision, thus greatly diversifying the breadth of applications. Since then, a host of applications for DNA, RNA and proteins analysis using both biological and solid-state nanopores have been demonstrated (4, 13, 14). Compared to their organic counterparts, solid-state nanopores were expected to emerge as an essential component of any practical nanopore-based instrumentation due to the size control, increased robustness of the membrane, and their natural propensity for integration with wafer-scale technologies, including CMOS and microfluidics (15, 16). Yet, this prospect is significantly hindered due to the constraints and limitations imposed by ion beam sculpting and transmission electron microscopy-based drilling, which, to this date, remain the only viable tools for achieving nanopores fabrication with dimensional control at the 1-nm scale. The complexity, low-throughput, and high-cost associated with these techniques restrict accessibility of the field to many researchers, greatly limit the productivity of the community, and prevent mass production of nanopore-based devices. Alternative nanofabrication strategies are therefore needed for the field to continue to thrive, and for the promised health-related applications to be successfully commercialized (including single-molecule DNA sequencing). Here, we introduce a fabrication method that is automated, simple, and low-cost, allowing nanopores to be created directly in aqueous solution with sub-nm precision, greatly facilitating use and opening up new paths toward nanomanufacturing of fluidics devices for a wide range of biotechnology applications.

## RESULTS AND DISCUSSION

We fabricate individual nanopores on thin insulating solid-state membranes directly in solution. A thin silicon nitride ($SiN_x$) membrane, supported by a silicon frame, is mounted in a liquid cell and separates two reservoirs containing an aqueous solution of 1M KCl. Ag/AgCl electrodes immersed on both sides of the membrane are connected to a custom-built resistive feedback current amplifier, which allow trans-membrane potentials of up to ±20V to be applied. The setup shown in Figure 1 is otherwise identical to what is commonly used for biomolecular detection (17), which greatly facilitates the transition to sensing experiments, eliminating further handling of membranes. See supporting information (SI) section 1 for more detail.

A single nanopore is fabricated by applying a constant potential difference, $\Delta V$, across a $t$ =10-nm or 30-nm thick $SiN_x$ membrane, to produce an electric field, $E=\Delta V/t$ in the dielectric membrane in the range of 0.4-1 V/nm (Figure 2a). At these high field strengths, a sustainable leakage current, $I_{leakage}$, is observed through the membrane, which remains otherwise insulating at low fields. $I_{leakage}$ rapidly increases with electric field strength, but is typically tens of nanoamperes for our operating conditions. We attribute the dominant conduction mechanism to a form of trap-assisted tunneling of electrons, supplied by ions in solution (18–21) (Figure 2b), since the membrane is too thick for significant direct tunnelling (18), and migration of impurities cannot produce lasting currents (22). Direct migration of electrolyte ions is also unlikely, since for a given electric field strength, a higher $I_{leakage}$ is observed in thicker membranes (Figure 2e). We provide additional discussion on the characteristics of the leakage current in the SI section 2.

We observe the creation of a single nanopore (i.e. fluidic channel spanning the membrane) by a sudden irreversible increase in $I_{leakage}$, which is attributed to the onset of ionic current (Figure 2f) due to a discrete dielectric breakdown event. As the current continues to increase, the nanopore further enlarges (Figure 2g). We use a feedback control mechanism to rapidly terminate the trans-membrane potential when the current exceeds a pre-determined threshold, $I_{cutoff}$. A threshold, set as $I_{cutoff}/I_{leakage}<1.2$, which is generally sufficient to terminate $\Delta V$ within ~0.1s of the breakdown event, can produce nanopores on the order of 2-nm in diameter as shown by the I-V curves in Figure 2h (see SI section 3 for additional results). In addition, following the nanopore fabrication event, we can continue to enlarge its size with sub-nm precision by applying moderate AC electric field square pulses in the range of ±0.2-0.3 V/nm, similar to Beamish *et al.* (23). This allows the nanopore size to be precisely tuned, for a particular sensing application, directly in neutral KCl solution.

**I-V Characteristics and Noise** – To infer the nanopore size upon fabrication, we measure its ionic conductance, $G$, and relate it to an effective diameter, $d$, assuming a cylindrical geometry and accounting for access resistance (24, 25), using:

$$G = \sigma \left[\frac{4t}{\pi d^2} + \frac{1}{d}\right]^{-1}$$

where $\sigma$ is the bulk conductivity of the solution. This method, practical for nanopores fabricated in liquids, provides a reasonable first order estimate of the pore size (25, 26) as confirmed by DNA translocations, and compares well with actual dimensions obtained from TEM images (see SI section 4 and 8). I-V curves are performed in a ±1 V window, where the leakage current can safely be ignored. Figure 2h reveals an ohmic electric response in 1M KCl. The majority of our nanopores exhibit linear I-V curves upon fabrication. The remaining nanopores that show signs of self-gating or rectification can be conditioned, by applying moderate electric field pulses (23), to slightly enlarge them until an ohmic behaviour is attained in high salt. Such I-V characteristics imply a relatively symmetric internal electric potential pore profile (27) which supports the symmetrical geometry with a uniform surface charge distribution assumed by our pore conductance model. Otherwise, one would expect significant rectification from multiple ≤1-nm fluidic paths or from a single narrow nano-crack of similar conductance, due to strong electrostatic double layer overlap. To further characterize the nanopores, we examined the noise in the ionic current by performing power spectral density measurements. Our fabrication method consistently produces nanopores with low-1/*f* noise levels, comparable to fully wetted TEM-drilled nanopores (see SI section 5)(28, 29). This may be attributed to the fact that nanopores are created directly in liquid rather than in vacuum. Thus far, we have successfully fabricated hundreds of individual nanopores ranging from 1 to 25-nm in size with comparable electrical characteristics that are stable for days.

**Dielectric Breakdown Mechanism –** In order that a single, well-defined nanopore be created each time, we postulate that the leakage current must be highly localized on the insulating membrane, since for conductive substrates (semiconductors or metals) anodic oxidation leads to an array of nanopores(30, 31). The leakage spot(s) must also modify the membrane at the nanoscale since an aqueous KCl solution at neutral pH is not an active etchant of $SiN_x$. To elucidate the mechanism leading to the formation of a nanopore, we investigate the fabrication process as a function of applied voltage, membrane thickness, electrolyte composition, concentration, and pH. Figure 3a shows the time-to-pore creation, $\tau$, as a function of the trans-membrane potential for 30-nm-thick membranes, in 1M KCl buffered at various pHs. Interestingly, $\tau$ scales exponentially with the applied voltage irrespective of other conditions, and can be as short as a few seconds. For a given voltage pH has a strong effect. As seen in Figure 3b, $\tau$ can be reduced by 1,000-fold when lowering the pH from 7 to 2. We have also observed that lower salt concentrations increase the fabrication time (data not shown). Overall, for a given fabrication condition $\tau$ is relatively consistent, though variations by a factor of 4 are common, and is uncorrelated with the size of the fabricated pore. Figure 3c shows $\tau$ for 10-nm-thick $SiN_x$ membranes, buffered at pH10 in various 1M Cl-based aqueous solutions. The fabrication time in these thinner membranes also decreases exponentially with potential, but the value required for forming a nanopore is now reduced by ~1/3 compared to 30-nm-thick membranes, irrespective of the different cations ($K^+$, $Na^+$, $Li^+$) tested. This observation indicates that the applied electric field in the membrane is the main driving force for initiating the fabrication of a single nanopore. Fields in the range of 0.4-1V/nm are close to the dielectric breakdown strength of low-stress $SiN_x$ films(19), and are key for intensifying the leakage current, which is thought to ultimately cause breakdown in thin insulating layers(32). The exponential dependence of $\tau$ on potential implies the same electric field dependency, which is reminiscent of the time-to-dielectric breakdown in gate dielectrics(32). According to the current understanding, dielectric breakdown mechanisms proceed as follows(32–34): (i) accumulation of charge traps (i.e. structural defects) by electric field-induced bond breakage or generated by charge injection from the anode or cathode, (ii) increasing up to a critical density forming a highly localized conductive path, and (iii) causing physical damage due to substantial power dissipation and the resultant heating. We propose that the process by which we fabricate a nanopore in solution is similar, though we control the damage to the nanoscale by limiting the localized leakage current, at the onset of the first, discrete breakdown event. Given the stochastic nature of the pore creation process, multiple simultaneous nanoscale breakdown events are unlikely, which ensures that ultimately a single nanopore is created. The process by which material is removed from the membrane remains unclear, but broken bonds could be chemically dissolved by the electrolyte or following a conversion to oxides/hydrides (35, 36), or sheared due to localized plasticity in the membrane. We explain the pH dependency on the fabrication time by the fact that breakdown at low pH is amplified by impact ionization producing an avalanche, due to the increased likelihood of hole injection or $H^+$ incorporation from the anode (see SI section 6 for more detail). To support the general character of nanofabrication by dielectric breakdown, we created nanopores in a different material (silicon dioxide) and present the data in the SI section 7.

**DNA Translocations –** We performed DNA translocation experiments to demonstrate that these nanopores can be leveraged for the benefit of single-molecule detection. Electrophoretically driven passage of a DNA molecule across a membrane is expected to transiently block the flow of ions in a manner that reflects the molecule length, size, charge and shape. The results using a ~6.4-nm-diameter pore, as estimated from conductance measurements, in a 10-nm thick $SiN_x$ membrane are shown in Figure 4. The scatter plot shows event duration and average current blockage of over 2,400 single-molecule translocations events of 5-kb dsDNA. The characteristic shape of the events is indistinguishable to data obtained on TEM-drilled nanopores (25, 37–39). The observed quantized current blockades strongly support the presence of a single nanopore spanning the membrane. Using dsDNA (~2.2 nm in diameter) as a molecular-sized ruler, the value of the single-level blockage events, $\Delta G = 7.4 \pm 0.9$nS, provides an effective pore diameter of 6.0 ±0.5-nm consistent with the size extracted from the pore conductance model (25). This result also suggests that the membrane thickness at the vicinity of the nanopores has not been significantly altered. We observed similar DNA signatures from most nanopores tested (>80% for N>20), and provide further discussion and additional translocation data in SI section 8.

**Conclusion –** Nanopore fabrication by controlled dielectric breakdown in solution represents a major reduction in complexity and cost over current fabrication methods, which will greatly facilitate accessibility to the field to many researchers, and provides a path to commercialize nanopore-based technologies. While we attribute the nanopore creation process to be an intrinsic property of the dielectric membrane, such that the nanopore can form anywhere on the surface, our current understanding strongly suggests that the position of the pore can be determined by locally controlling the electric field strength or the material dielectric strength. This could be achieved, for instance, by nanopatterning or locally thinning the membrane, by positioning of a nanoelectrode, or by confining the field to specific areas on the membrane via micro- or nanofluidic channel encapsulation (see SI section 9). The latter would also allow for the simple integration of independently addressable nanopores in an array format on a single chip.

## MATERIALS AND METHODS

**Dielectric Membranes** – Silicon Nitride ($SiN_x$) membranes used in our experiments are commercially available as transmission electron microscope (TEM) windows (Norcada product # NT005X and NT005Z). Each membrane is made of 10-nm or 30-nm thick low-stress (<250MPa) $SiN_x$, deposited on 200-μm thick lightly doped silicon (Si) substrate by low-pressure chemical vapour deposition (LPCVD). A 50-μm × 50-μm window on the backside of the Si substrate is opened by a KOH anisotropic chemical etch. The absence of pre-existing structural damages (e.g. pinholes, nano-cracks) is inferred by the fact that no current (<pA) is measured across a membrane at low voltages (<±1V) prior to nanopore fabrication. Silicon dioxide membranes were also purchased from TEMWindows (product# SO100-A20Q33). Note that we have also successfully fabricated nanopores on $SiN_x$ membranes purchased from TEMWindows, and on custom fabricated $SiN_x$ membranes.

**Instrumentation and Data Acquisition** – A schematic of the experimental setup is shown in Figure 1. A silicon chip with an intact silicon nitride membrane is sandwiched between two silicone gaskets (shown in purple on the figure). It is then positioned between the two electrolyte reservoirs in a PTFE or a PEEK (polytetrafluoroethylene) (polyether ether ketone) fluidic cell. The two reservoirs filled with liquid electrolyte are electrically connected to a current amplifier by two Ag/AgCl electrodes. The entire system is encapsulated in a grounded faraday cage to isolate from electromagnetic interference. Data acquisition and measurement automation were performed using custom-designed LabVIEW software controlling a National Instruments USB-6351 or PXIe-6366 DAQ card. The value of the trans-membrane potential is set the DAQ card. Leakage current is digitized at 250 kHz and the signal is filtered at 10 Hz. When a current exceed a pre-set threshold, the voltage bias is immediately ceased by the software (response time is ~100 ms). Ionic current signal during DNA translocations are recorded using an Axopatch 200B with a 4-pole Bessel filter set at 100kHz, with at 250kHz sampling rate. Data analysis was carried out using custom-designed LabVIEW software to measure the duration and depth of each current blockade events.

**DNA Studies** – We performed DNA translocation studies, using dsDNA fragments of 100bp, 5kbp, 10kbp purchased from Fermantas (NoLimits products) in 1M KCl pH8 or in 3.6M LiCl pH8 at a final concentration of 10μg/mL. Lambda DNA (48.5kbp) purchased from NewEngland BioLabs was also used.

**ACKNOWLEDGEMENTS** – This work was supported by the Natural Sciences and Engineering Research Council of Canada, the Canada Foundation for Innovation, and Ontario Network of Excellence. The authors would like to thank Y. Liu for aid in TEM imaging and L. Andrzejewski for valuable technical support.

## REFERENCES

1. Venkatesan BM, Bashir R (2011) Nanopore sensors for nucleic acid analysis. *Nat Nanotechnol* 6:615–24.

2. Dekker C (2007) Solid-state nanopores. *Nat Nanotechnol* 2:209–15.

3. Branton D et al. (2008) The potential and challenges of nanopore sequencing. *Nat Biotechnol* 26:1146–53.

4. Kasianowicz JJ, Robertson JWF, Chan ER, Reiner JE, Stanford VM (2008) Nanoscopic porous sensors. *Annu Rev Anal Chem (Palo Alto Calif)* 1:737–66.

5. Bezrukov SM, Vodyanoy I, Parsegian VA (1994) Counting polymers moving through a single ion channel. *Nature* 370:279–81.

6. Kasianowicz JJ, Brandin E, Branton D, Deamer DW (1996) Characterization of individual polynucleotide molecules using a membrane channel. *Proc Natl Acad Sci U S A* 93:13770–3.

7. Li J et al. (2001) Ion-beam sculpting at nanometre length scales. *Nature* 412:166–9.

8. Storm AJ, Chen JH, Ling XS, Zandbergen HW, Dekker C (2003) Fabrication of solid-state nanopores with single-nanometre precision. *Nat Mater* 2:537–40.

9. Storm a. J, Chen JH, Ling XS, Zandbergen HW, Dekker C (2005) Electron-beam-induced deformations of SiO[sub 2] nanostructures. *J Appl Phys* 98:014307.

10. Kuan AT, Golovchenko JA (2012) Nanometer-thin solid-state nanopores by cold ion beam sculpting. *Appl Phys Lett* 100:213104–2131044.

11. Russo CJ, Golovchenko JA (2012) Atom-by-atom nucleation and growth of graphene nanopores. *Proc Natl Acad Sci U S A* 109:5953–7.

12. Yang J et al. (2011) Rapid and precise scanning helium ion microscope milling of solid-state nanopores for biomolecule detection. *Nanotechnology* 22:285310.

13. Miles BN et al. (2013) Single molecule sensing with solid-state nanopores: novel materials, methods, and applications. *Chem Soc Rev* 42:15–28.

14. Oukhaled A, Bacri L, Pastoriza-Gallego M, Betton J-M, Pelta J (2012) Sensing Proteins through Nanopores: Fundamental to Applications. *ACS Chem Biol* 7:1935–1949.

15. Rosenstein JK, Wanunu M, Merchant CA, Drndic M, Shepard KL (2012) Integrated nanopore sensing platform with sub-microsecond temporal resolution. *Nat Methods* 9:487–492.

16. Jain T, Guerrero RJS, Aguilar CA, Karnik R (2013) Integration of solid-state nanopores in microfluidic networks via transfer printing of suspended membranes. *Anal Chem* 85:3871–8.

17. Tabard-Cossa V (2013) *Engineered Nanopores for Bioanalytical Applications - Chapter 3: Instrumentation for Low-Noise High-Bandwidth Nanopore Recording* eds Edel JB, Albrecht T (Elsevier Inc.)1st edition.

18. Frenkel J (1938) On Pre-Breakdown Phenomena in Insulators and Electronic Semi-Conductors. *Phys Rev* 54:647–648.

19. Habermehl S, Apodaca RT, Kaplar RJ (2009) On dielectric breakdown in silicon-rich silicon nitride thin films. *Appl Phys Lett* 94:012905.

20. Jeong DS, Hwang CS (2005) Tunneling-assisted Poole-Frenkel conduction mechanism in HfO[sub 2] thin films. *J Appl Phys* 98:113701.

21. Kimura M, Ohmi T (1996) Conduction mechanism and origin of stress-induced leakage current in thin silicon dioxide films. *J Appl Phys* 80:6360.

22. Lee S, An R, Hunt AJ (2010) Liquid glass electrodes for nanofluidics. *Nat Nanotechnol* 5:412–6.

23. Beamish E, Kwok H, Tabard-Cossa V, Godin M (2012) Precise control of the size and noise of solid-state nanopores using high electric fields. *Nanotechnology* 23:405301.

24. Vodyanoy I, Bezrukov SM (1992) Sizing of an ion pore by access resistance measurements. *Biophys J* 62:10–1.

25. Kowalczyk SW, Grosberg AY, Rabin Y, Dekker C (2011) Modeling the conductance and DNA blockade of solid-state nanopores. *Nanotechnology* 22:315101.

26. Frament CM, Dwyer JR (2012) Conductance-Based Determination of Solid-State Nanopore Size and Shape: An Exploration of Performance Limits. *J Phys Chem C* 116:23315–23321.

27. Kosińska ID (2006) How the asymmetry of internal potential influences the shape of I-V characteristic of nanochannels. *J Chem Phys* 124:244707.

28. Tabard-Cossa V, Trivedi D, Wiggin M, Jetha NN, Marziali A (2007) Noise analysis and reduction in solid-state nanopores. *Nanotechnology* 18:305505.

29. Smeets RMM, Keyser UF, Dekker NH, Dekker C (2008) Noise in solid-state nanopores. *Proc Natl Acad Sci U S A* 105:417–21.

30. Thompson GE, Wood GC (1981) Porous anodic film formation on aluminium. *Nature* 290:230–232.

31. Létant SE, Hart BR, Van Buuren AW, Terminello LJ (2003) Functionalized silicon membranes for selective bio-organism capture. *Nat Mater* 2:391–5.

32. Lombardo S et al. (2005) Dielectric breakdown mechanisms in gate oxides. *J Appl Phys* 98:121301.

33. McPherson JW, Mogul HC (1998) Underlying physics of the thermochemical E model in describing low-field time-dependent dielectric breakdown in SiO[sub 2] thin films. *J Appl Phys* 84:1513.

34. DiMaria DJ, Cartier E, Arnold D (1993) Impact ionization, trap creation, degradation, and breakdown in silicon dioxide films on silicon. *J Appl Phys* 73:3367.

35. Liu H, Steigerwald ML, Nuckolls C (2009) Electrical double layer catalyzed wet-etching of silicon dioxide. *J Am Chem Soc* 131:17034–5.

36. Jamasb S, Collins S, Smith RL (1998) A physical model for drift in pH ISFETs. *Sensors Actuators B Chem* 49:146–155.

37. Chen P et al. (2004) Probing Single DNA Molecule Transport Using Fabricated Nanopores. *Nano Lett* 4:2293–2298.

38. Fologea D, Brandin E, Uplinger J, Branton D, Li J (2007) DNA conformation and base number simultaneously determined in a nanopore. *Electrophoresis* 28:3186–92.

39. Li J, Gershow M, Stein D, Brandin E, Golovchenko J a (2003) DNA molecules and configurations in a solid-state nanopore microscope. *Nat Mater* 2:611–5.

**Supplementary Information:** This file contains Supplementary Text and Data 1-9, Supplementary Figures 1-11 , Tables 1-2, and additional references.

**FIGURES**

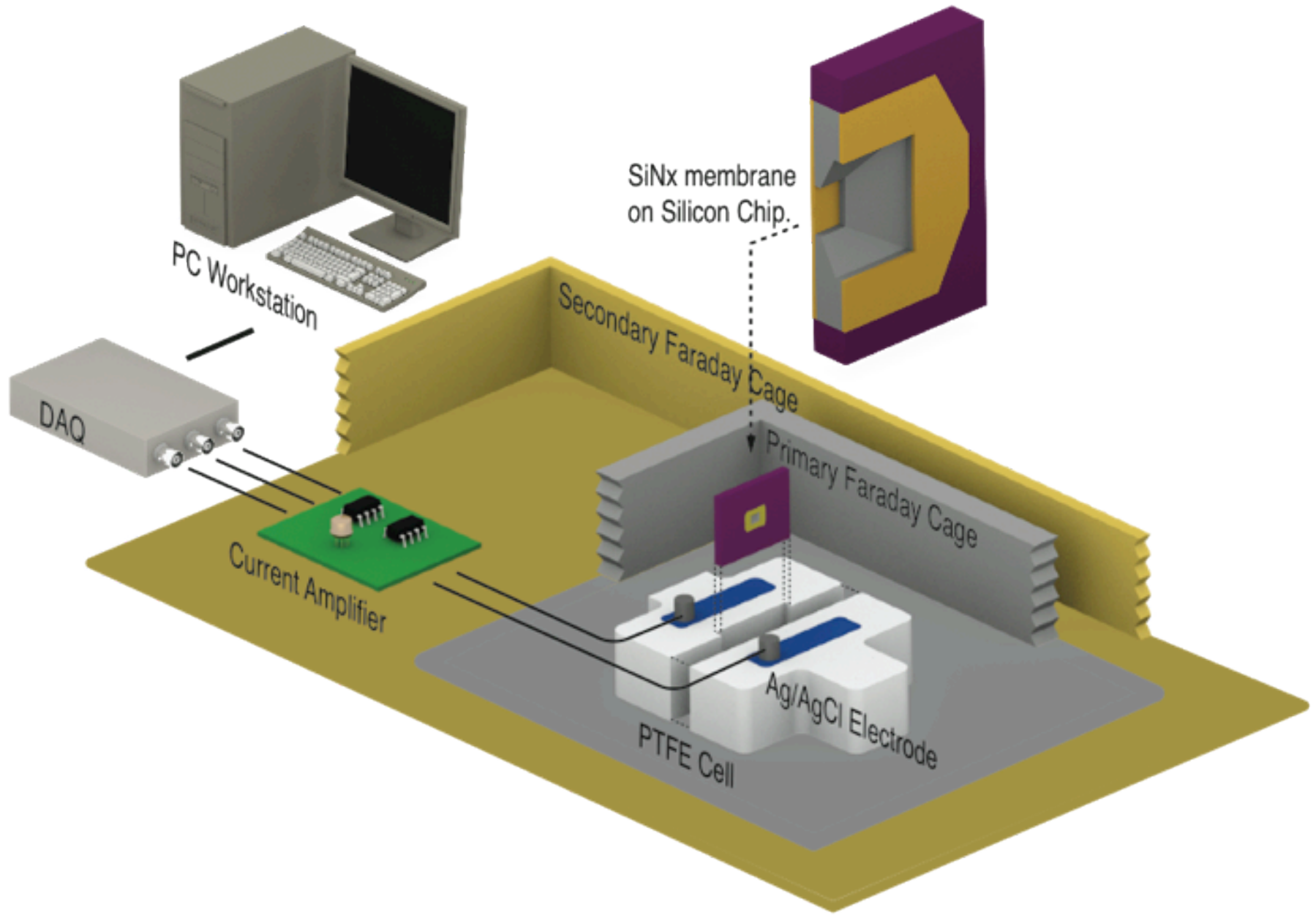

**Figure 1: Schematic of the experimental setup used to fabricate nanopores directly in aqueous solutions. A computer-controlled custom current amplifier is used to apply voltages up to ±20V and measure the current with sub-nA sensitivity from one of the two Ag/AgCl electrodes positioned on either sides of the membrane. It is noteworthy to realize that this experimental setup is identical - with the exception of the custom current amplifier replacing the commonly used Axopatch 200B (Molecular Devices) - to the instrumentation used to study DNA or proteins translocation through nanopores.**

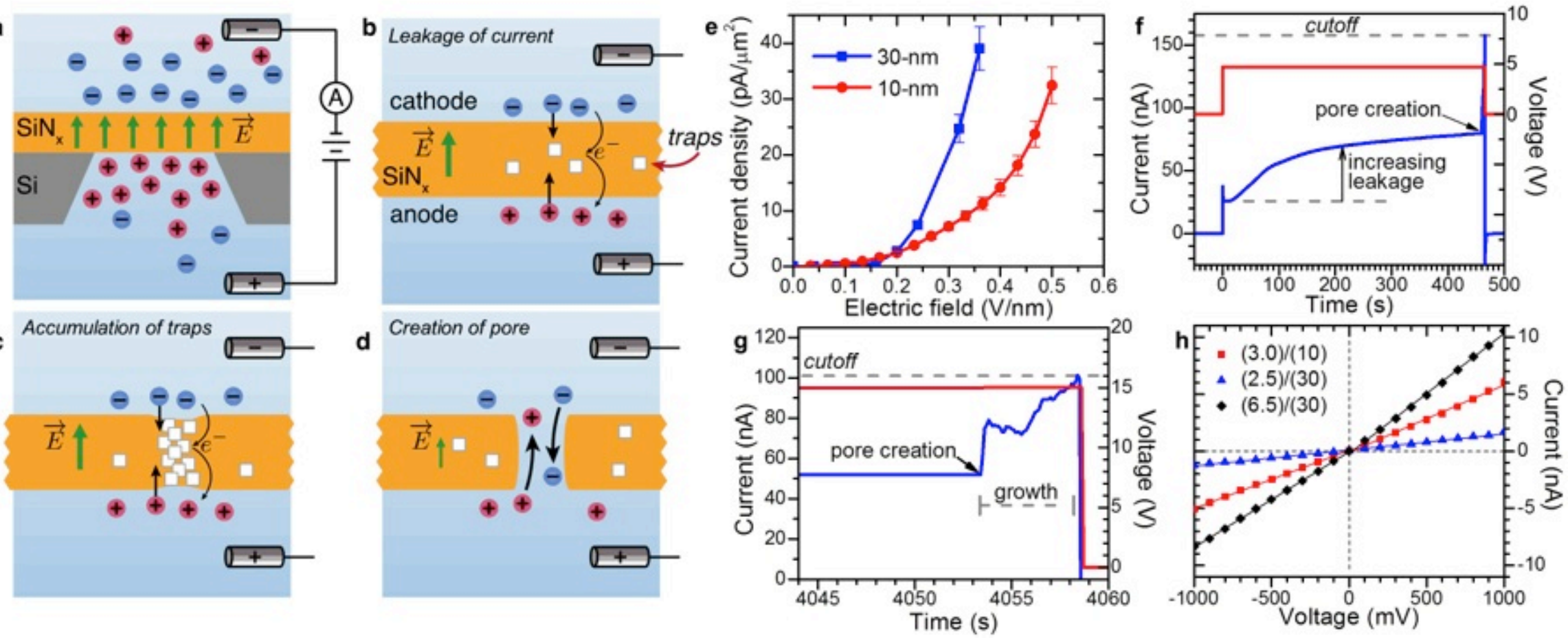

**Figure 2: a) Application of a trans-membrane potential generates an electric inside the $SiN_x$, and charges the interfaces with opposite ions. b) Leakage current through the membrane follows a trap-assisted tunneling mechanism. Free charges (electrons or holes) can be produced by redox reactions at the surface or by field ionization of incorporated ions. The number of available charged traps (structural defects) sets the magnitude of the observed leakage current. c) Accumulation of charge traps produced by electric field-induced bond breakage or energetic charges carries leads to a highly localized conductive path, and a discrete dielectric breakdown event. d) A nanopore is formed following dissolution of the defects. e) Leakage current density for $SiN_x$ membranes (50-μm × 50-μm). The leakage current is fully reversible and stable, unless high fields are sustained, see SI section 2. A larger current is observed on thicker membranes since the number of charge traps (defects) per unit area is greater as their number in the material increases with volume. f) Leakage current during nanopore fabrication process. Experiment performed at 5 V, on a 10-nm-thick $SiN_x$ membrane, in 1M KCl at pH13.5. Pore created is ~5-nm (18 nS). The slowly increase leakage current, following the capacitive spike, is a result of the accumulation of traps in the membrane. g) Experiment performed at 15 V, on a 30-nm-thick $SiN_x$ membrane, in 1M KCl pH10. The nanopore is allowed to grow until a pre-determined threshold current is reached, at which point the trans-membrane potential is turned off. The observed current fluctuations at the onset of pore formation are attributed to significant low-frequency noise at this voltage. Pore created is ~3-nm (2.9 nS). h) Current-to-voltage curves for 3 independent nanopores fabricated on different membranes. The legend indicates the (pore diameter)/(membrane thickness) in nm. Measurements performed in 1M KCl pH8, with an Axopatch 200B.**

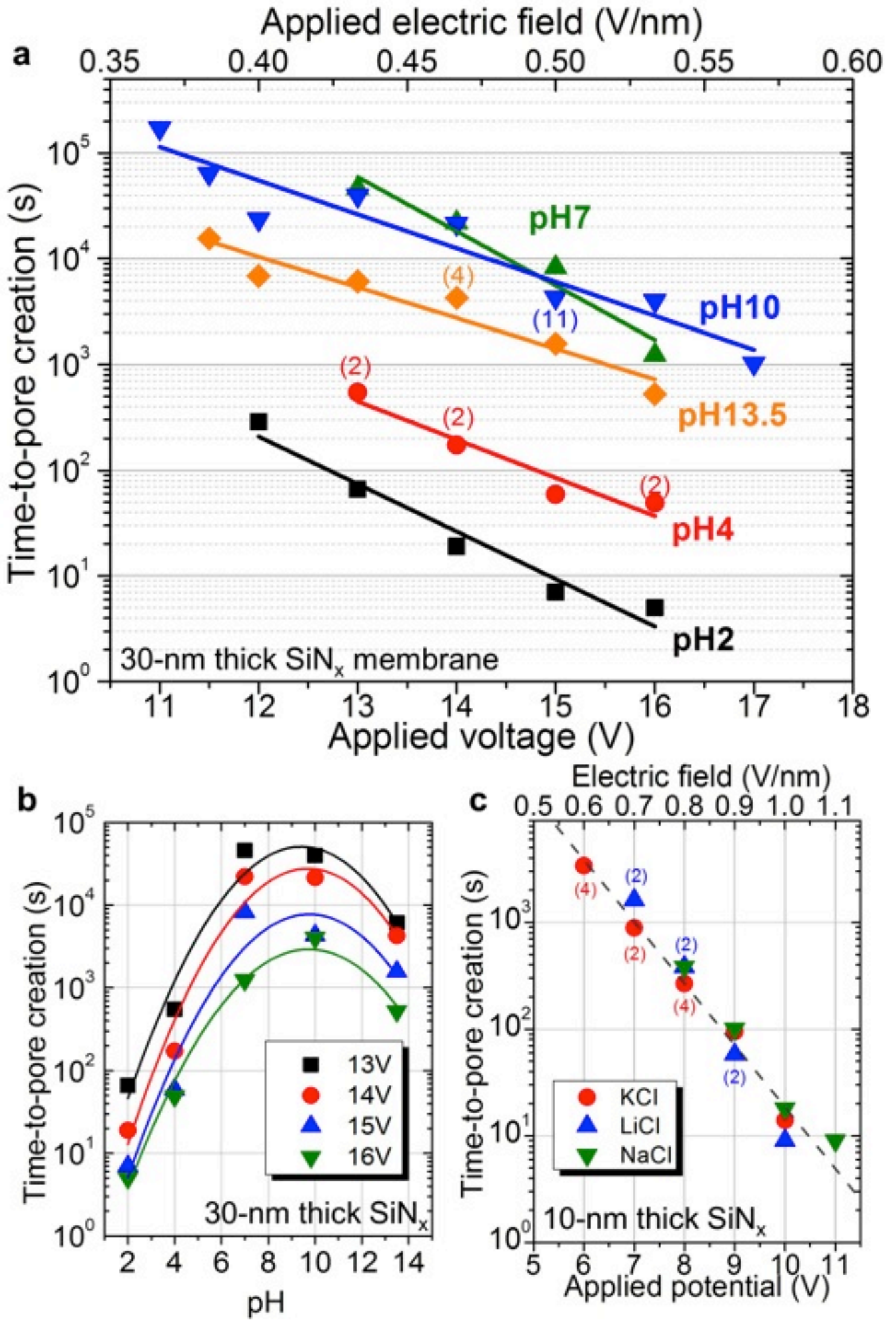

**Figure 3: a) Semi-log plot of fabrication time of individual nanopores created in 30-nm-thick $SiN_x$ membranes in 1M KCl buffered as indicated, versus the applied voltage and the calculated applied electric field. The number of separate nanopores each data point is averaged over is indicated in parentheses. The vast majority of nanopores plotted are sub-5-nm in size (i.e. <7 nS). b) Semi-log plot of fabrication time versus pH for the data plotted in a). c) Semi-log plot of fabrication time of individual nanopores created in 10-nm-thick $SiN_x$ membranes in 1M Cl-based electrolyte buffered at pH 10 for different cationic species versus the applied voltage and the calculated applied electric field. All nanopores plotted are sub-5-nm in size (i.e. <20 nS).**

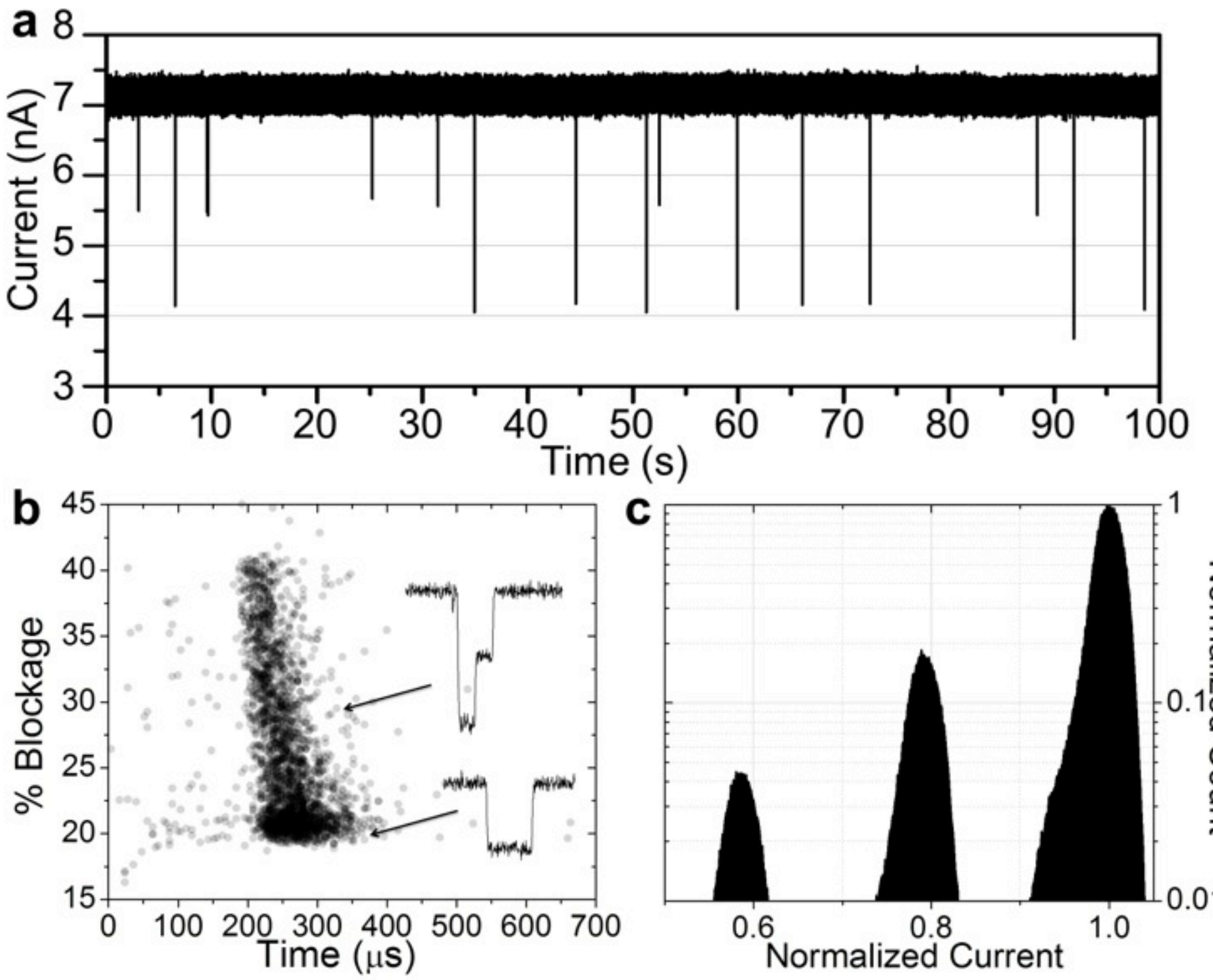

**Figure 4: a) Ionic current trace showing multiple DNA translocation events through a ~6.4-nm pore in a 10-nm-thick $SiN_x$ membrane. Experiments performed with 10ug/mL of 5-kb DNA fragments in 3.6M LiCl pH8, at 200mV using an Axopatch 200B. Data sampled at 250 kHz and low-pass filtered at 100 kHz. b) Scatter plot of the normalized average current blockade (0% presenting a fully opened pore, and 100% a fully blocked pore) versus the total translocation time of a single-molecule event. Each data point represents a single DNA translocation event. The majority of the events are unfolded. There are very few anomalously long events, indicating weak DNA-pore interactions. The inset shows ionic current signatures of two single-molecule translocation events, passing in a linear and partially folded conformation. c) Histogram of the current level revealing the expected quantization of the amplitude of current blockades. Quantized levels corresponding to zero, one, two dsDNA strands in the nanopore are clearly observed.**

# Supplementary Information

*"Nanopore Fabrication by Controlled Dielectric Breakdown"*

Harold Kwok, Kyle Briggs, and Vincent Tabard-Cossa

*Center for Interdisciplinary Nanophysics, Department of Physics, University of Ottawa, 150 Louis Pasteur, Ottawa, Ontario K1N 6N5, Canada*

## S1. Procedures, and Experimental Setup

For fabricating a nanopore, a custom current amplifier is employed to measure the leakage current while applying a voltage bias of up to ±20V. Current is measured at the ground Ag/AgCl electrode with pA sensitivity. The circuit relies on a simple operation-amplifier circuit (as shown in figure S1) to read and control voltage and current. Current signal is digitized by a data acquisition circuit, and is continuously being fed to a computer. Current is monitored in real time at a frequency of 10Hz. When the current exceeds a pre-set threshold, voltage bias is ceased immediately.

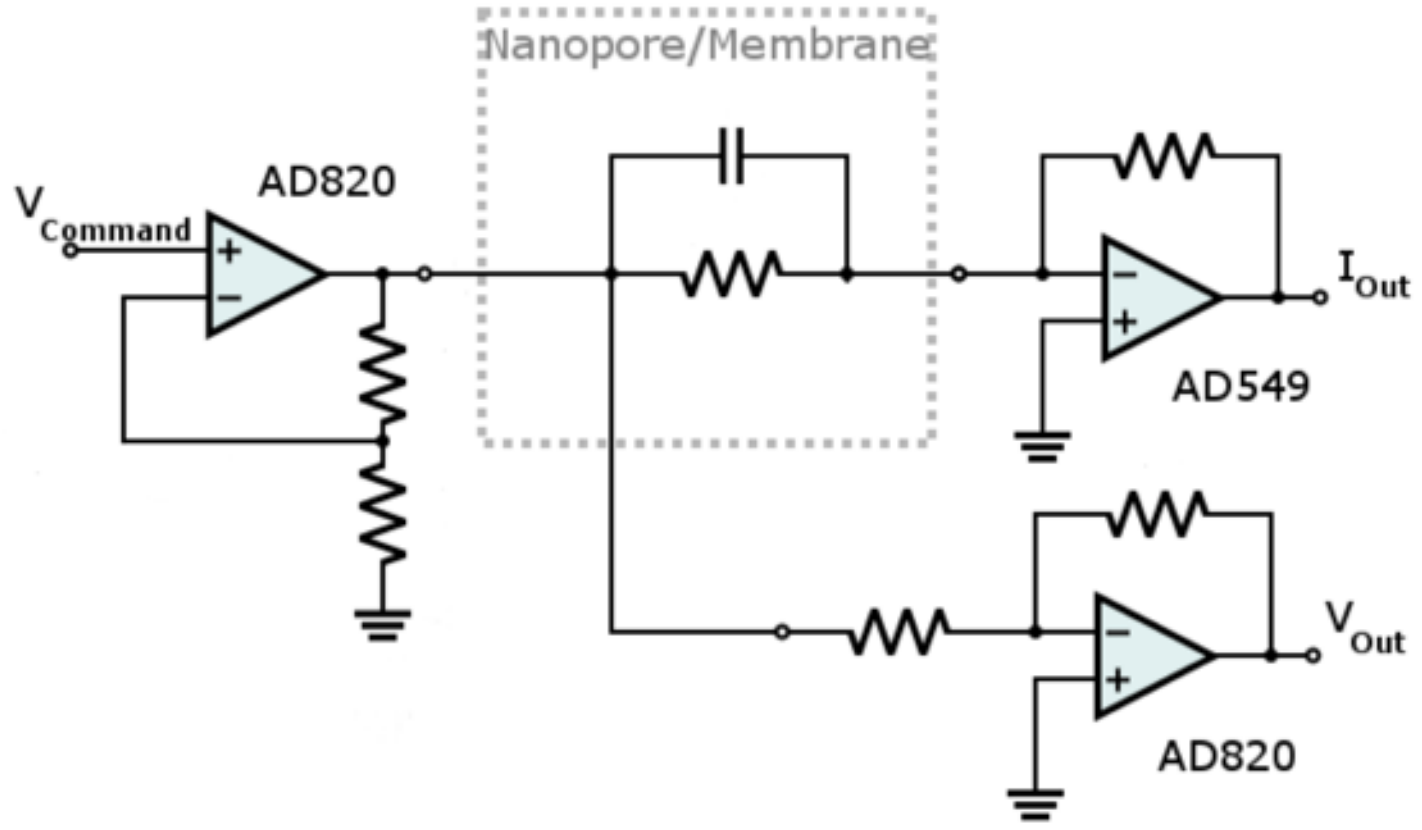

**Figure S1: Schematic of the custom-built current amplifier. Op-amps used are AD820 and AD549. All op-amps are powered by a ±20V voltage source. The circuit takes in a command voltage between ±10V from a computer controlled DAQ card, which is amplified to ±20V, and sets the trans-membrane potential. Current is measured with a transimpedance amplifier topology (AD549 with a 5MΩ feedback resistor), and the signal digitized by a DAQ card. The applied trans-membrane potential is also measured. The signal is scaled by 1/10 before being digitized by a DAQ card.**

After the creation of a nanopore, the custom current amplifier is replaced by a commercial amplifier, Axopatch 200B (Molecular Devices) – Figure S2. Its special architecture allows for lower noise at higher bandwidth recording of ionic current, but can only apply trans-membrane potentials up to ±1V. Noise characterisation, *I-V* response and DNA translocation events are recorded at 250kHz sampling rate by this instrument operating at the voltage clamp mode with a 100kHz 4-pole Bessel low-pass filter.

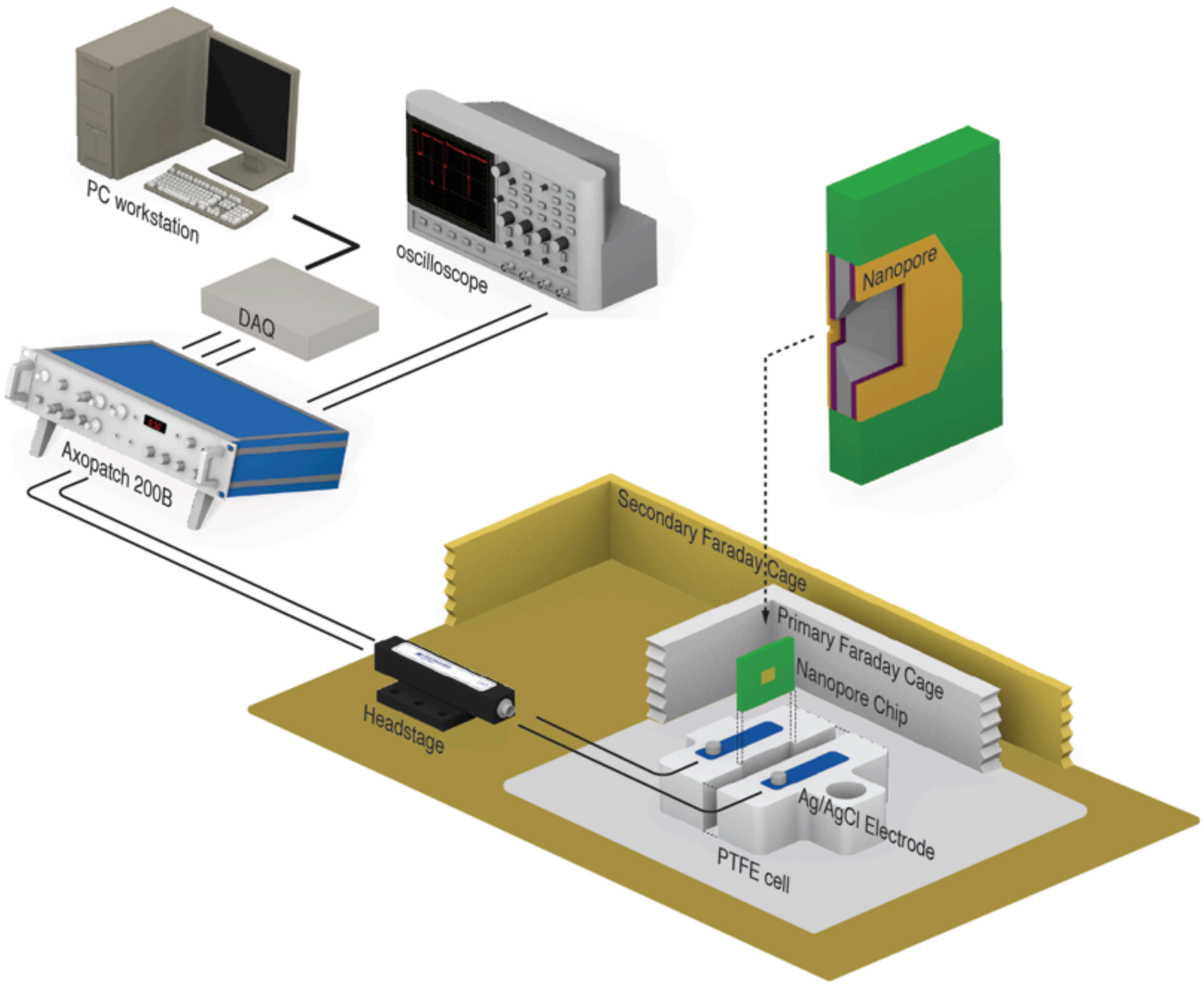

**Figure S2: Schematic of the experimental setup used for DNA translocation and noise characterization. The custom current amplifier is replaced with an Axopatch 200B for low-noise high-bandwidth recordings.**

## S2. Leakage Current

Leakage current is measured across an intact silicon nitride membrane as a function of trans-membrane potential, while immersed in an electrolyte solution (Figure 1a). At the onset of the applied voltage, a current spike from capacitive current persists for a few hundred milliseconds (proportional to the chip capacitance), followed by a slowly varying leakage current (see Figure 1f), which usually gradually increases. We note that this is not always the case, and under long-term fabrication conditions (>hours), leakage current has also been observed to be stable or decrease for different set of experimental conditions as shown in the Figure S3. This behavior is expected to be highly material dependent. The observed leakage current is attributed to a trap-assisted tunneling of electrons through the membrane. We argue that the slow increase corresponds to the stress induced leakage current (SILC)(1–4), which is caused by the accumulation of charge traps within the dielectric material. The measured leakage current shown in Figure 1e was recorded after 4s of an applied voltage, to remove the effect of the capacitive current spike, and avoid observing SILC. The latter generally prevents data points above 0.4V/nm to be plotted on Figure 1e since the leakage current does not reach a stable state.

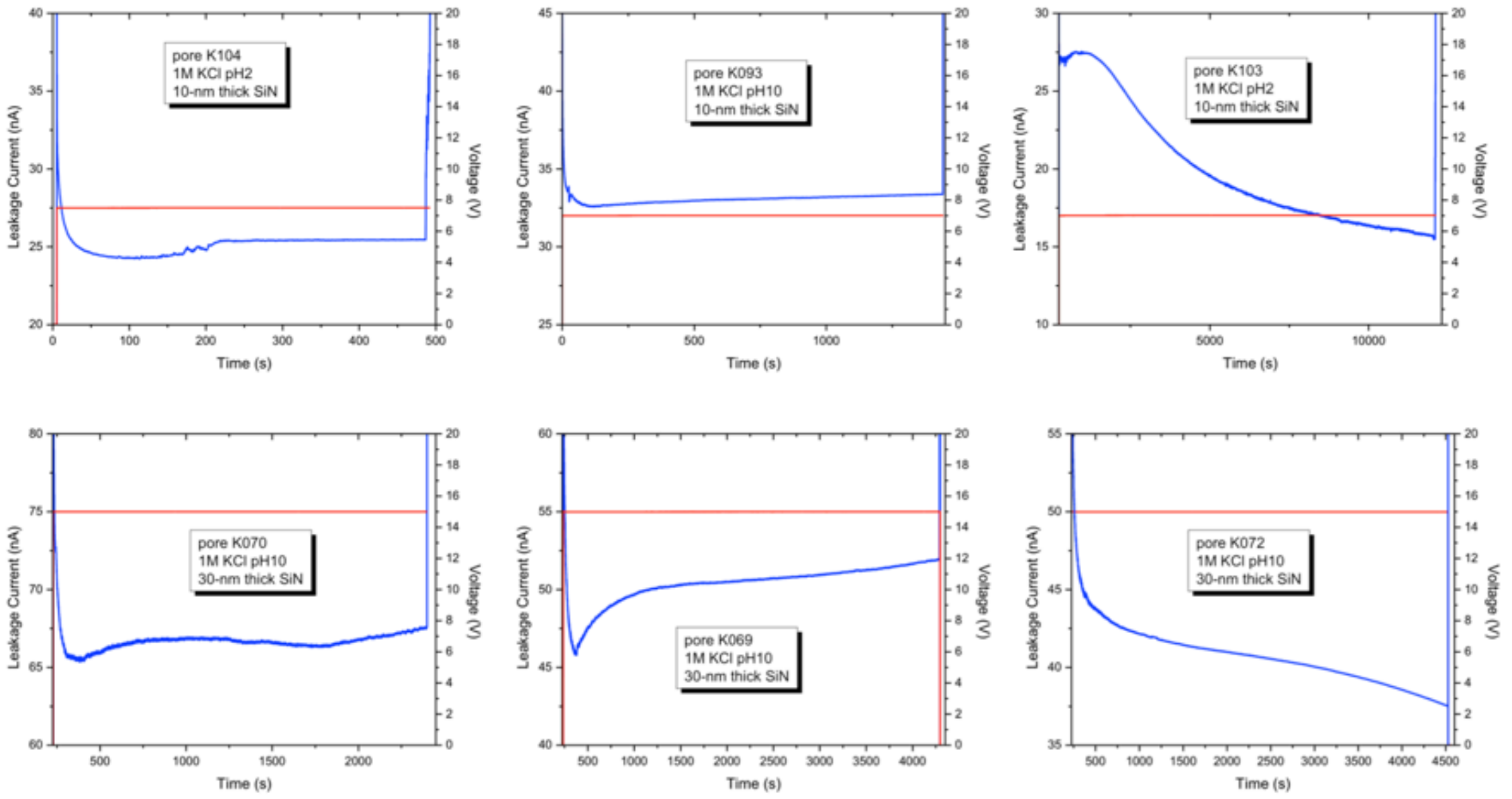

**Figure S3: representative leakage current traces during nanopore fabrication for both 10-nm and 30-nm thick SiN membranes. While the leakage current often slowly changes with time, the onset of pore creation characterized by an abrupt change in current is always clearly identifiable.**

## S3. IV curves of ~2-nm nanopores

We present, in Figure S4, I-V curves for 8 nanopores with effective diameters ranging between 1.4-nm and 2.4-nm, as calculated from our conductance-based model accounting for access resistance, to demonstrate that nanopores on the order of ~2-nm can be fabricated. The frequency response of our control software was set to 10Hz (equivalent to 100ms response time to stop the current from increasing after crossing the threshold) and the cutoff current was relatively tight (20% increase from the leakage current baseline) – Note that these settings are not particularly severe.

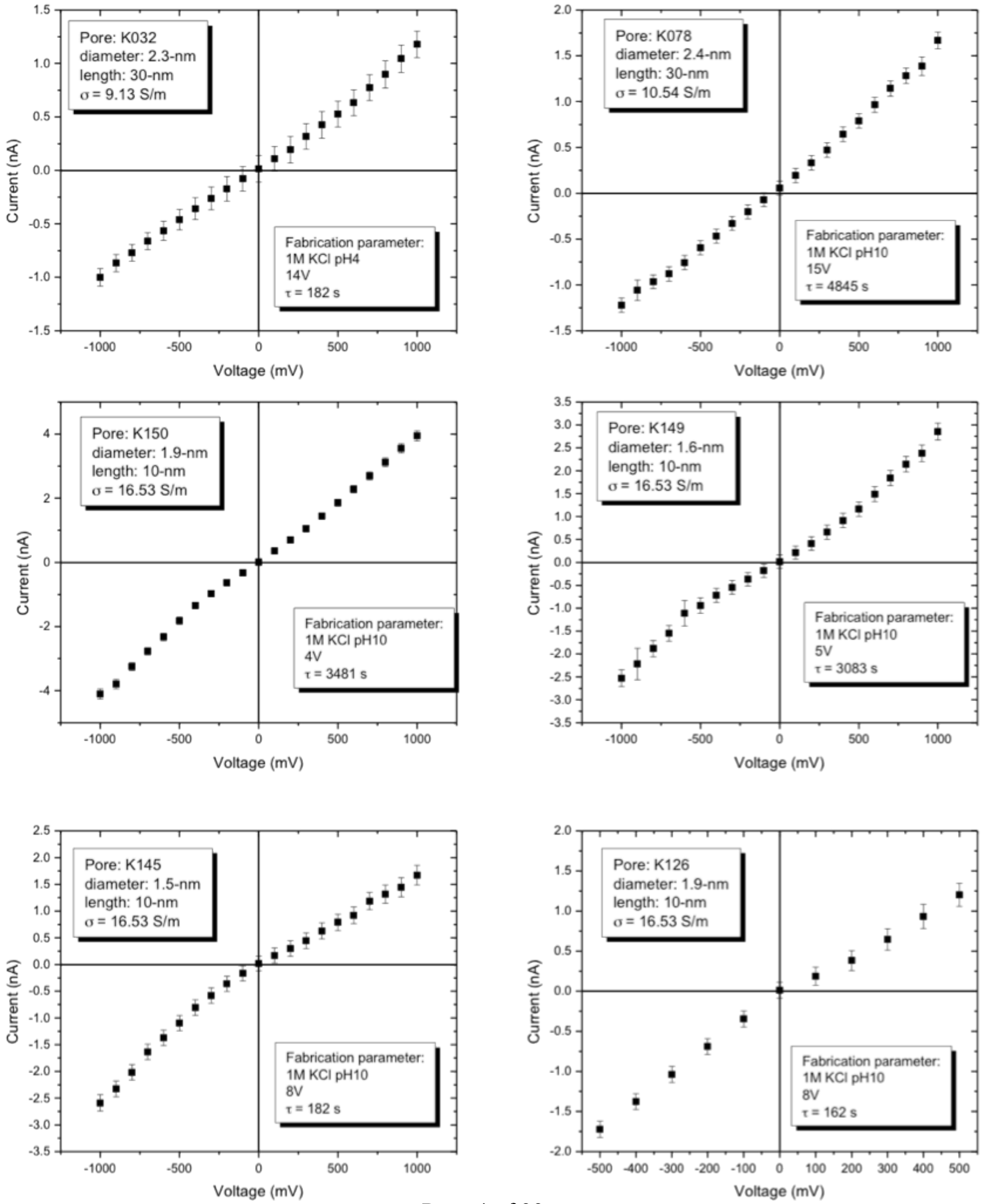

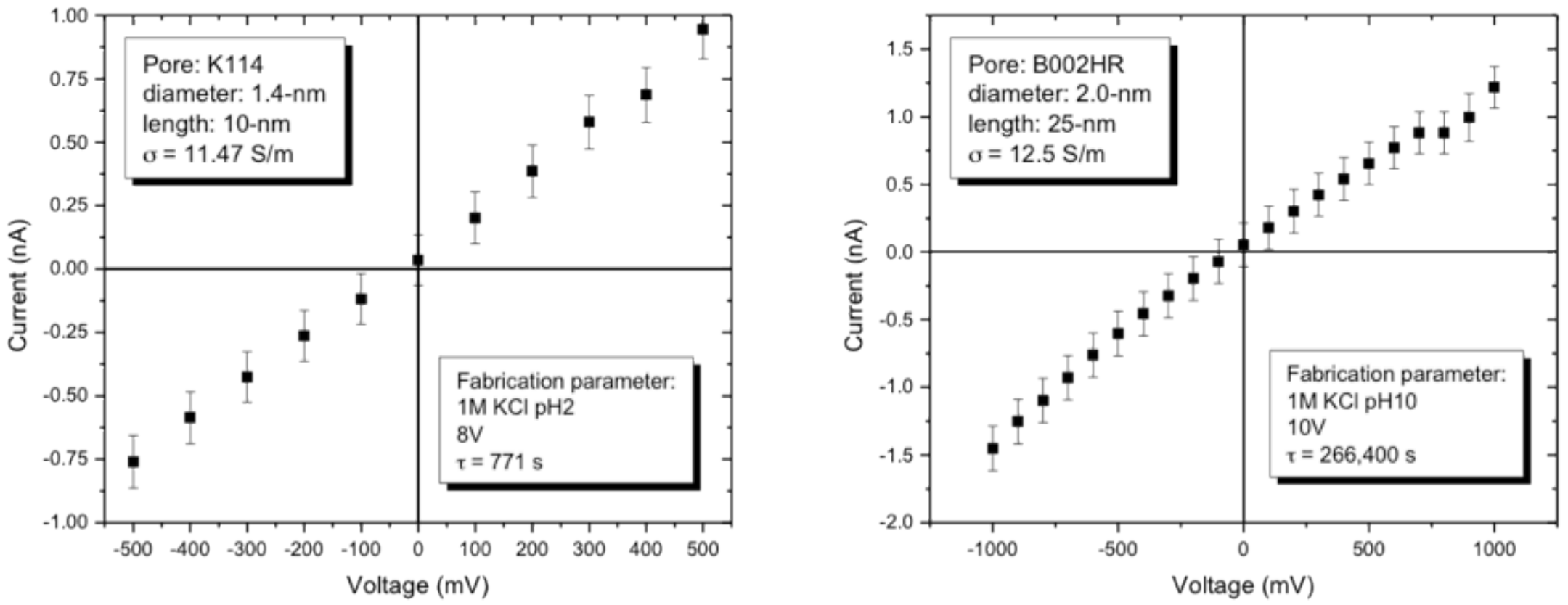

**Figure S4: I-V curves of 8 independent nanopores fabricated on 10-nm and 30-nm thick SiN membranes, using different fabrication conditions**.

We note that while approximations of our conductance-based model can affect the accuracy of the values extracted for the effective nanopore diameter (e.g. deviation from exact cylindrical geometry from pore to pore), data included in the manuscript and supplementary information sections generally suggest a <1-nm error. Remarkably, this level of control for the automated nanofabrication of solid-state nanopores in thin dielectric membranes surpasses what is currently possible with the state-of-the art electron or ion beam drilling techniques.

## S4. TEM Imaging

Imaging a nanopore fabricated by dielectric breakdown in solution under a transmission electron microscope can be a laborious task. Indeed, we suspect the nanopore creation process to be an intrinsic property of the dielectric membrane, such that the nanopore can form anywhere on the surface. The commercial membranes employed in this work are 50-μm×50-μm. Looking for a sub-10-nm feature on a $25\times10^2$-$\mu m^2$ area is time-consuming and somewhat difficult since other debris (e.g. salt residues) can be present. Our attempts to locate and image a pore on these membranes failed.

In order to obtain TEM images of pores we therefore used a different type of membrane, with a reduced window size, graciously provided by Stratos Genomics. These custom devices have a 38-nm thick silicon nitride membrane <10-μm in size. Figure S5 shows TEM images of two independent nanopores fabricated by dielectric breakdown in solution on these custom membranes. We have observed the following three important characteristics from the nanopore images obtained thus far:

1. A top view of pores confirms a circular pore opening, which was assumed by our conductance-based model.
2. Comprehensive exploration of the membrane surface revealed the presence of a single nanopore. No other partially formed pores or unusual membrane features have been observed thus far.
3. The dimensions obtained for the TEM image are in good agreement with the size extracted from the conductance-based model, especially for pores >5-nm. Due to increased importance of surface effects (i.e. small variations of pore wall profile and surface charge density), we expect the accuracy of the conductance-based model to be lower for sub-5-nm pores in 1M KCl, particularly for thicker membranes.

These initial observations tend to indicate that the calculated pore diameter extracted from our conductance-based model may somewhat underestimate the actual diameter, though it is possible that the cleaning procedure (a few hours immersed in warm DI water) to remove salt residues may alter the pore dimensions. Nevertheless, these results support the existence of a single fluidic channel spanning the membrane, as otherwise the observed conductance would be significantly larger than what is expected from the measured TEM size of a single nanopore.

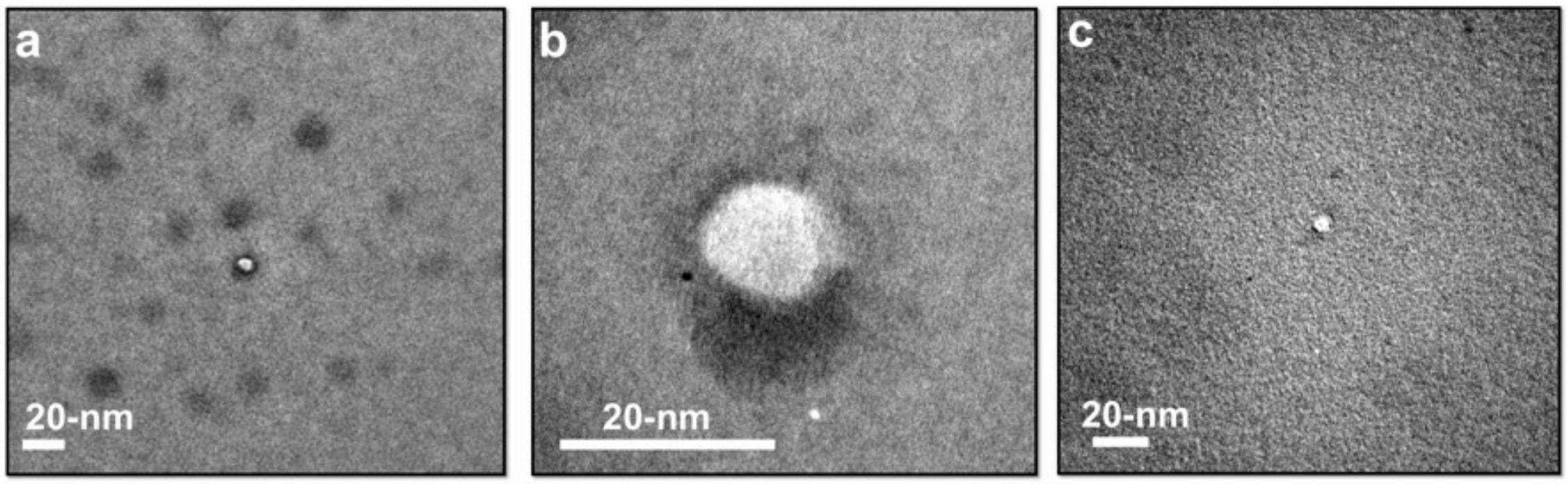

**Figure S5: TEM images of nanopores fabricated by dielectric breakdown in solution. a) image of a nanopore with an estimated pore diameter extracted from conductance-based model of ~11-nm. b) Image of the same pore taken at higher magnification. Pore size measured by TEM is ~14-nm. c) Image of different nanopore on a distinct membrane with an estimated pore diameter extracted from conductance-based model of ~3-nm. Pore size measured by TEM is ~5-nm. Dark spots in the image in a) are observed on intact membranes before being immersed in solution and subjected to high electric fields. These spots are believed to be silicon rich regions, which are due to the particular LPCVD process used. Such dark spots are not observed on the commercial membranes from Norcada. The small bright spot below the pore image in b) is a defective area on the CCD, and is found on every image. Note that it is cropped from the image a).**

## S5. Noise Characterization

Noise is characterized by recording a time series of ionic current at a constant voltage below ±1V. We examined noise in the ionic current flowing through our fabricated nanopores, by performing power spectral density (PSD) measurements to determine their noise level. These measurements were performed using Axopatch 200B (Molecular Devices) as the current amplifier with the 4-pole Bessel filter set at 100kHz. Figure S6 shows the PSD of different membranes containing fabricated nanopores under an applied trans-membrane potential of 200mV, in 1M KCl pH8. Remarkably, the PSD for these nanopores fabricated *in solution* reveals low $1/f$ noise level, comparable to the best TEM-drilled nanopores(5, 6), and in some instances ultra-low $1/f$ noise level can be attained, which are comparable to biological pores(5). We noticed that, similarly to TEM-drilled pores, nanopores created by dielectric breakdown do exhibit variability in their low-frequency noise level, though low $1/f$ noise level can be obtained with very high yield (>80%). Nevertheless, in the event of a high noise nanopore, the low-frequency noise can be reduced following a conditioning procedure similar to Beamish et al.(7). The fact that we can reliably produce nanopores with low $1/f$ noise may be attributed to the fact that our nanopores are created directly in aqueous solution and are never exposed to air, thus eliminating wetting issues, and minimizing the likelihood of nanobubbles trapped at the vicinity of a pore(8). In addition, we speculate that the nanopore's inner wall surface can be chemically modified by the fabrication process, which may contribute to an improved noise performance.

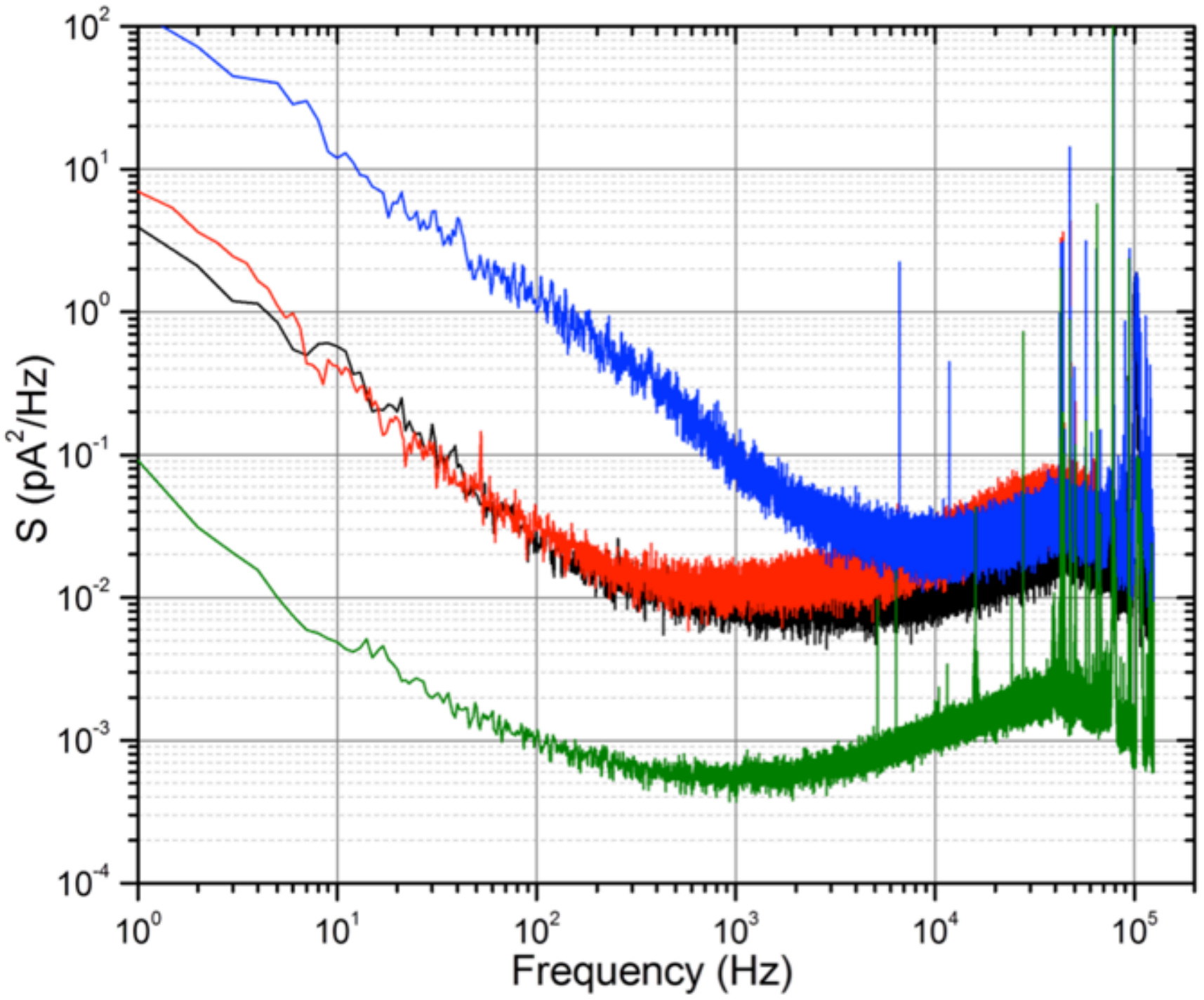

**Figure S6: Power spectrum densities of the ionic current of four nanopore fabricated by controlled dielectric breakdown, recorded by an Axopatch 200B at 200mV. Data sampled at 250kHz, and low-pass filtered at 100kHz by a 4-pole Bessel filter. *Blue curve* - pore is 6-nm in diameter, created at pH2 in 1M KCl. *Red curve* - pore is 6-nm in diameter, created at pH2 in 1M KCl. *Black curve* - pore is 6-nm in diameter, created at pH2 in 1M KCl. *Green curve* - pore is 12-nm in diameter, created at pH13.5 in 1M KCl. The chip is covered with polydimethylsiloxane (PDMS) to minimize noises from membrane capacitance(5). At this time, there does not yet appear to be any obvious relationship between noise levels and fabrication conditions.**

## S6. Time-to-pore creation versus Trans-membrane potential and Electric Field strength.

The fabrication of a nanopore in a dielectric membrane is similar in nature to the time-dependent dielectric breakdown (TDDB) process in solid-state devices. It is therefore interesting to show the time required to create a single nanopore in relation to the applied electric field and voltage. Here we re-plot some of the data showed in Figure 2. We tested two membrane thickness values ($t$ = 10-nm, 30-nm), at various pH and applied voltages, $V$. The applied electric field is given by $E = V/t$. Figure S7 shows an approximate linear relationship in semi-log scale for the time-to-pore fabrication versus the applied voltage or applied field. This trend is reminiscent to dielectric breakdown models for solid-state devices(2, 9). However, fabrication times for the two different membrane thicknesses studied do not always follow the same behaviour. In alkaline conditions, 10-nm and 30-nm thick membrane collapse on a continuous curve as a function of applied electric field, indicating that the field strength dominates the fabrication process. In acidic conditions however, 10-nm membranes require a much higher field to create a nanopore for a given fabrication time. We speculate that the strong pH dependence observed in 30-nm thick membranes is a consequence of the greater thickness of the membrane, enabling impact ionization effects to take place. In 30-nm thick membrane, we therefore argue that breakdown at low pH is amplified by impact ionization-induced avalanche, due to the increased likelihood of $H^+$ incorporation or hole injection from the anode side of the membrane(10). Nevertheless, the possibility of pH and voltage driven chemical and electrochemical reactions participating in the pore fabrication process should not be ignored and could also be playing a role in the observed results.

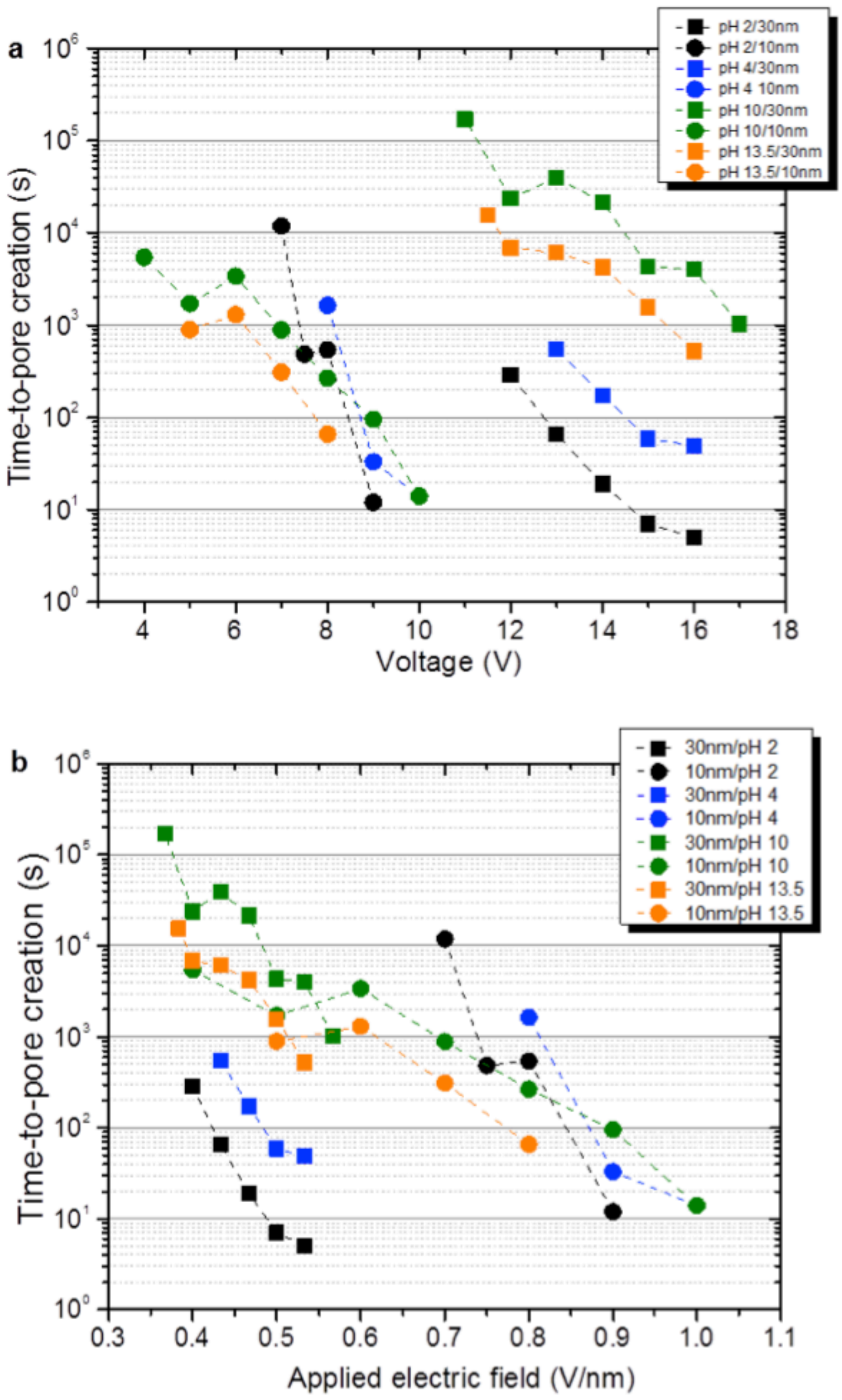

**Figure S7: Time-to-pore creation a nanopore plotted in a log scale against a) applied voltage, b) calculated electric field strength, for 10-nm and 30-nm thick silicon nitride membranes, in 1M KCl at various pHs. Some of this data is plotted in Figure 2. Some data points are averaged over multiple creation events.**

## S7. Fabrication on $SiO_2$ Membranes.

We fabricated nanopores in 20-nm thick $SiO_2$ membranes purchased from SIMPore (TEMWindows – product# SO100-A20Q33). A weaker electric field strength was needed to fabricate a nanopore in $SiO_2$ in a given time compared to SiN, which we attribute to the reduced dielectric constant, and is consistent with dielectric breakdown mechanism for the pore formation. Figure S8 shows the nanopore creation event during a 2 s pulse at 7.5V and continued enlargement at 8V, as well as the resulting I-V curve of the nanopore, showing an effective diameter of 6.2-nm. We successfully confirmed the ability of the nanopore to detect DNA by studying the translocation of 10-kpb dsDNA molecules.

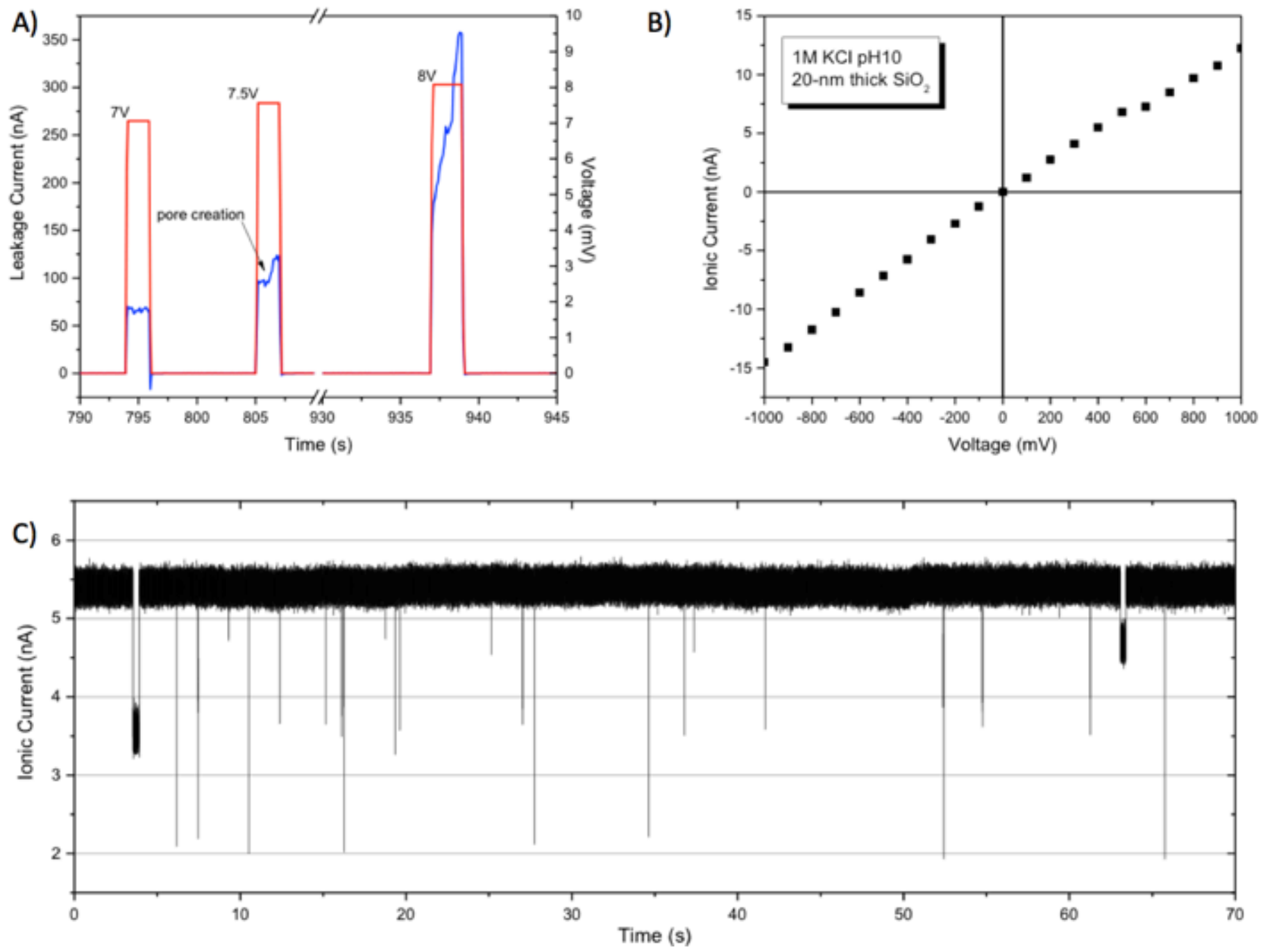

**Figure S8: A) Leakage current versus time for different strength of applied voltage pulses. Pore creation happens during the 2s pulse at 7.5V, and is enlarged during the subsequent 8V pulse. B) I-V curve of the resulting nanopore, with an effective diameter of 6.2-nm. C) Ionic current trace at 400mV in 1M LiCl pH8, showing characteristics blockages from translocating 10kpb dsDNA molecules.**

## S8. DNA Translocation

The ability to detect single molecule events, such as DNA translocation, is an essential feature of a nanopore-based sensing device. To illustrate that nanopores fabricated by our method are compatible for single bio-molecule sensing applications, we performed a systematic investigation of 18 independent nanopores fabricated in a wide range of experimental conditions in 10-nm thick silicon nitride membranes. 11 nanopores were between 4.5-nm and 6.5-nm in effective size, 4 nanopores were between 7-nm and 9-nm in effective size, and 3 nanopores were between 10-nm and 25-nm in effective size. These nanopores were systematically tested by injecting 5-kb dsDNA fragments (Fermentas, NoLimits) on one side of the membrane in 3.6M LiCl pH8 at a final concentration of 10μg/mL. LiCl was chosen to maximize the current blockades and increase translocation times(11). Amongst the 18 nanopores tested, 15 successfully showed DNA translocation events, while 9 of them detected >1,000 events before clogging[*]. In other DNA translocation studies, we have also demonstrated detection of 100-pb DNA fragments, 10-kb DNA fragments, Lamda DNA molecules, and 10-nm and 30-nm thick $SiN_x$ membranes, and 20-nm thick $SiO_2$ membranes. The yield is comparable to if not greater than TEM-drilled nanopores (based on our experience and from discussions with other groups). However, we should point out that these statistics were obtained without any optimization in the fabrication conditions, and we therefore believe that major improvement can be achieved.

Below, we present additional DNA translocation results of a 6.5-nm pore in a 10-nm thick silicon nitride membrane (fabrication parameters: 7V, 1M KCl pH2, $\tau$=11,837s), in 3.6M LiCl buffered at pH8, at five different voltages ranging between 200mV and 1000mV. Current traces showing some of these events are shown in Figure S9.

[*] We note that clogged pores can often be unclogged by following a process similar to Beamish et al.(7), though at the expense of an increase in size, which can nevertheless be minimized to a few nanometers.

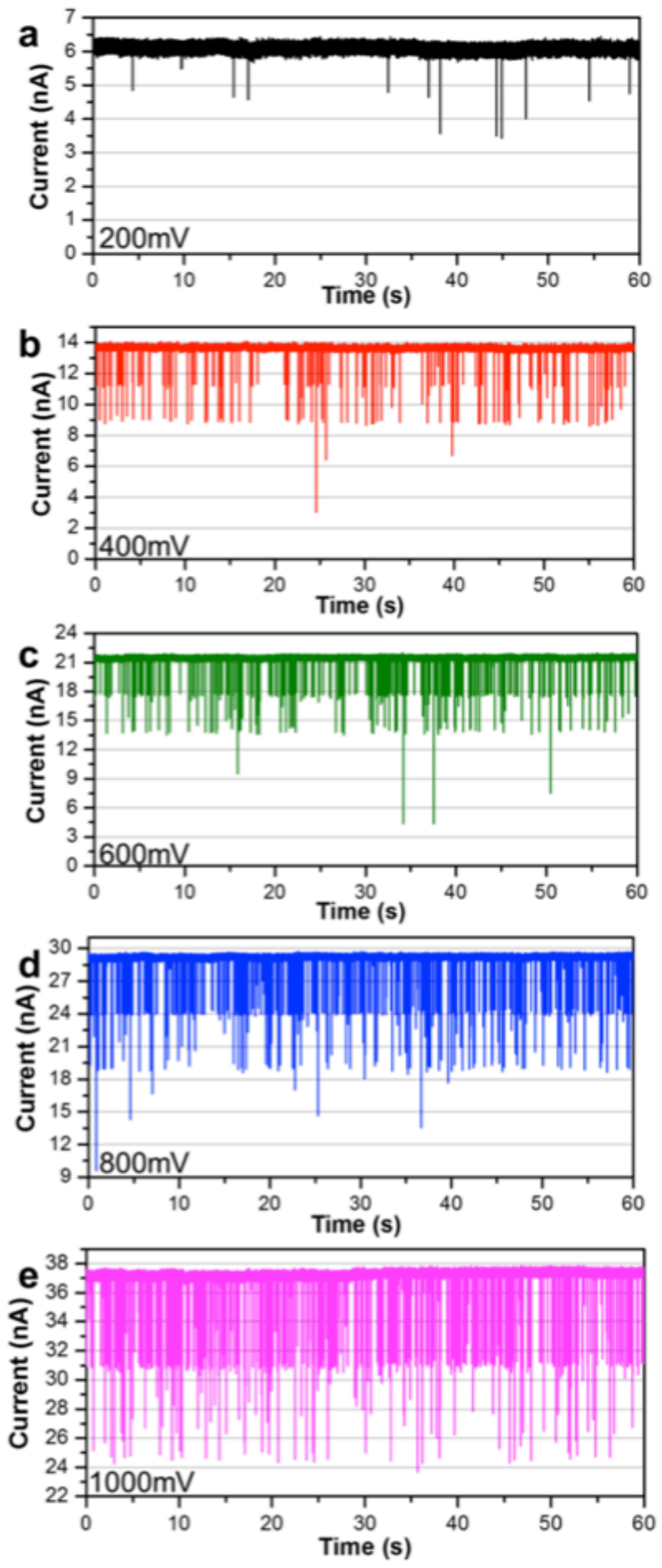

**Figure S9: Current versus time of 5-kbp dsDNA translocation events for a 6.5-nm pore in a 10-nm membrane in pH 8 3.6M LiCl. Data recording at 250kHz sampling rate, and low-pass filtered by a 4-pole Bessel filter set at 100kHz. a) Data at 200mV was recording when the nanopore had an effective size of 6.0-nm. The pore partially clogged after a few hundreds events. A few pulses of moderate electric fields (0.2 V/nm) were used to unclogged it, which also increased its effective size to 6.5-nm. b-e) Data at 400mV, 600mV, 800mV and 1000mV was subsequently recorded without flushing out or adding new DNA molecules.**

**Expected Conductance Blockades –** These current traces, with observed blockades showing multiple quantized ionic current levels, are characteristic of long negatively charged polymers electrophoretically forced through a nanopore and compare qualitatively well with published results obtained on TEM-drilled nanopores(12–14). The interpretation that the observed current blockades are indeed due to dsDNA translocation through nanopores, fabricated by our method, is supported by both the increase in current blockage amplitude and the decrease in the mean translocation time with increasing trans-membrane potential across the nanopore (see also Figures S10 and S11). The expected current blockage from a dsDNA molecule translocating through a nanopore will depend on the particular geometry of the nanopore, and the electrolyte composition(15). Translocation data obtained from nanopore fabricated by controlled dielectric breakdown compares very well with results obtained from TEM-drilled pores. Similar to what has been previously done for TEM-drilled pores, our results can be rationalized by assuming a cylindrical nanopore geometry, from which we can approximate the expected conductance blockade, $\Delta G$, from a single strand of dsDNA spanning the length of the pore as(12, 16):

$$\Delta G = \frac{\sigma \pi d_{DNA}^2}{4L} \quad \text{(S1)}$$

where $\sigma$ is the solution conductivity, and $L$ the pore length. This simple expression based on a cylindrical model of the pore is frequently used in the nanopore literature(17) and allows $L$ to be determined without a priori knowledge of the pore diameter. In the limit of high salt concentration (3.6M LiCl, used in this case) and small nanopores (<10-nm), edge effects become less important due to both increased effective shielding of the electrolyte and reduced relative importance of the access resistance as compared to pore resistance (see Figure 1 of Kowalczyk et al.(17)). We therefore argue that the approximations of equation (S1) affect only slightly the accuracy of our results (see the Table1 below, comparing the various values obtained with different models).

Solving equation (S1) assuming a diameter of B-form double-stranded DNA of 2.2 nm, and taking the effective pore length as the nominal membrane thickness (10-nm), we obtain $\Delta G$ = 6.3 nS for the experimental conditions used in Figure S12-S13 (conductivity of 16.45 S/m). Our experimental results, of $\Delta G$ = 6.2 ± 0.2nS, for the various trans-membrane potentials tested are therefore consistent with a first-order cylindrical approximation of the nanopore geometry.

**Calculation of the effective pore length** – DNA translocation results can be used to verify that the membrane thickness has not been thinned in a noticeable way during the nanopore fabrication process. We can calculate the effective pore length at the pore location using equation (S1) and compare it to the nominal membrane thickness.

We have performed an in-depth analysis of 8 independent DNA translocation experiments on 10-nm thick silicon nitride membranes, totally ~30,000 events showing unfolded conductance blockades, and using a range of fabrication conditions. Table 1 and Table 2 summarize our results, which show an average effective pore length, $L$ = 9.0 ± 0.3nm. This value is within the tolerance range of the manufacturer of these commercial membranes, quoted as a thickness of 10 ± 1nm (product #: Norcada NT005Z).

**Table 1: Translocations were performed in 3.6M LiCl pH 8, with 5kbp dsDNA, except pore K097 which used 10kpb dsDNA. Nominal membrane thickness is 10-nm.**

| Pore ID | $V(mV)$ | $G(nS)$ | $\Delta G(nS)$ | $L\ (nm)$ | $d\ (nm)$ | $d_{simple}$ $(nm)$ | # of events |
|---|---|---|---|---|---|---|---|
| **K093** *(pore shown in text)* | 200 | $35.4 \pm 0.6$ | $7.4 \pm 0.9$ | $8 \pm 1$ | $6.0 \pm 0.5$ | 4.8 | 2490 |
| | 400 | $33.8 \pm 0.3$ | $6.6 \pm 0.6$ | $9 \pm 1$ | $6.1 \pm 0.5$ | 5.0 | 190 |
| | 600 | $36.4 \pm 0.3$ | $7.1 \pm 0.5$ | $9 \pm 1$ | $6.2 \pm 0.5$ | 5.0 | 549 |
| **K097** | 200 | $23.0 \pm 0.4$ | $8.7 \pm 0.6$ | $7.2 \pm 0.9$ | $4.3 \pm 0.3$ | 3.6 | 277 |
| | 400 | $25.1 \pm 0.3$ | $8.9 \pm 0.4$ | $7.0 \pm 0.8$ | $4.5 \pm 0.3$ | 3.7 | 166 |
| | 600 | $25.3 \pm 0.1$ | $8.3 \pm 0.1$ | $7.5 \pm 0.8$ | $4.7 \pm 0.3$ | 3.8 | 346 |
| | 1000 | $27.1 \pm 0.2$ | $8.8 \pm 0.3$ | $7.1 \pm 0.7$ | $4.8 \pm 0.4$ | 3.9 | 384 |
| **K099** | 200 | $25.3 \pm 0.6$ | $5.6 \pm 0.9$ | $11 \pm 2$ | $5.5 \pm 0.5$ | 4.7 | 722 |
| | 400 | $27.7 \pm 0.3$ | $6.0 \pm 0.4$ | $10 \pm 1$ | $5.6 \pm 0.4$ | 4.7 | 2172 |
| | 600 | $28. \pm 0.2$ | $6.4 \pm 0.3$ | $10 \pm 1$ | $5.6 \pm 0.4$ | 4.7 | 1724 |
| **K103** *(pore shown in supp. info)* | 200 | $29.8 \pm 0.6$ | $5.7 \pm 0.9$ | $11 \pm 2$ | $6.0 \pm 0.6$ | 5.0 | 161 |
| | 400 | $34.1 \pm 0.1$ | $6.0 \pm 0.2$ | $10 \pm 1$ | $6.4 \pm 0.5$ | 5.2 | 2867 |
| | 600 | $35.8 \pm 0.4$ | $6.4 \pm 0.5$ | $10 \pm 1$ | $6.4 \pm 0.5$ | 5.2 | 4117 |
| | 800 | $36.5 \pm 0.2$ | $6.3 \pm 0.2$ | $10 \pm 1$ | $6.5 \pm 0.5$ | 5.3 | 3408 |
| | 1000 | $37.3 \pm 0.1$ | $6.4 \pm 0.1$ | $10 \pm 1$ | $6.6 \pm 0.5$ | 5.3 | 747 |
| **K104** | 200 | $32.3 \pm 0.6$ | $7.0 \pm 0.9$ | $9 \pm 2$ | $5.8 \pm 0.5$ | 4.7 | 120 |
| | 400 | $32.4 \pm 0.3$ | $7.1 \pm 0.4$ | $9 \pm 1$ | $5.8 \pm 0.4$ | 4.7 | 879 |
| | 600 | $32.9 \pm 0.3$ | $7.1 \pm 0.4$ | $9 \pm 1$ | $5.8 \pm 0.4$ | 4.7 | 1247 |
| | 800 | $33.1 \pm 0.2$ | $7.3 \pm 0.5$ | $9 \pm 1$ | $5.8 \pm 0.4$ | 4.7 | 1070 |
| | 1000 | $33.9 \pm 0.2$ | $7.1 \pm 0.4$ | $9 \pm 1$ | $5.9 \pm 0.4$ | 4.8 | 1176 |
| **K106** | 200 | $43.7 \pm 0.9$ | $6 \pm 1$ | $11 \pm 3$ | $7.6 \pm 0.8$ | 6.1 | 175 |
| | 1000 | $46.6 \pm 0.2$ | $5.7 \pm 0.2$ | $11 \pm 1$ | $7.9 \pm 0.6$ | 6.3 | 83 |
| **K123** | 400 | $32.9 \pm 0.2$ | $8.1 \pm 0.4$ | $7.7 \pm 0.8$ | $5.5 \pm 0.4$ | 4.4 | 1119 |
| | 600 | $33.3 \pm 0.3$ | $8.4 \pm 0.4$ | $7.4 \pm 0.8$ | $5.5 \pm 0.4$ | 4.4 | 839 |
| | 800 | $34.6 \pm 0.3$ | $8.1 \pm 0.3$ | $7.7 \pm 0.8$ | $5.7 \pm 0.4$ | 4.5 | 2798 |
| | 1000 | $35.1 \pm 0.2$ | $8.2 \pm 0.2$ | $7.6 \pm 0.8$ | $5.7 \pm 0.4$ | 4.6 | 586 |
| **K124** | 200 | $32.9 \pm 0.6$ | $9 \pm 1$ | $7 \pm 1$ | $5.4 \pm 0.5$ | 4.3 | 496 |
| | 400 | $31.4 \pm 0.3$ | $8.2 \pm 0.5$ | $7.6 \pm 0.9$ | $5.4 \pm 0.4$ | 4.3 | 1451 |
| | 600 | $33.9 \pm 0.2$ | $8.6 \pm 0.3$ | $7.3 \pm 0.8$ | $5.5 \pm 0.4$ | 4.4 | 849 |
| | 800 | $34.6 \pm 0.2$ | $8.4 \pm 0.3$ | $7.4 \pm 0.8$ | $5.6 \pm 0.4$ | 4.5 | 1203 |
| | 1000 | $35.6 \pm 0.1$ | $8.1 \pm 0.1$ | $7.7 \pm 0.8$ | $5.8 \pm 0.4$ | 4.6 | 69 |

**Note that:**

- **K093 was fabricated in pH 10 1M KCl at -9V for 360s, and conditioned at ±3V AC in pH 8 3.6M LiCl for 800s.**
- **K097 was fabricated in pH 10 1M KCl at -8V for 258s and conditioned at ±3V AC in pH 8 3.6M LiCl for 3900s.**
- **K099 was fabricated in pH 2 1M KCl at -8V for 300s and conditioned at ±3V AC in pH 8 3.6M LiCl for 4634s.**
- **K103 was fabricated in pH 2 1M KCl at -7V for 11837s and conditioned at ±4V AC in pH 8 3.6M LiCl for 4500s.**
- **K104 was fabricated in pH 2 1M KCl at -7.5V for 482s and conditioned at ±3V AC in pH 8 3.6M LiCl for 8884s.**
- **K106 was fabricated in pH 13.5 1M KCl at -8V for 66s and conditioned at ±3V AC in pH 8 3.6M LiCl for 2360s.**
- **K123 was fabricated in pH 10 1M NaCl at -10V for 18s and conditioned at ±3V AC in pH 8 3.6M LiCl for 1270s.**
- **K124 was fabricated in pH 10 1M NaCl at -9V for 101s and conditioned at ±3V AC in pH 8 3.6M LiCl for 540s**

Error on the calculated membrane thickness was determined as follow:

$$\delta \mathrm{L} = \frac{A_{DNA}}{\Delta G}\sqrt{\delta\sigma^2 + \frac{\sigma^2 \delta\Delta G^2}{\Delta G^2}}$$

where error on the solution conductivity was estimated at 10%, and error on the conductance blockade is taken as the square root of the sums of the squares of the HWHMs histograms of the first blocked state and open pore state.

We then use this value of effective membrane thickness $L$ to refine our pore diameter $d$ using open pore conductance $G = \sigma\left(\frac{4L}{\pi d^2} + \frac{1}{d}\right)^{-1}$, obtaining

$$d = \frac{G}{2\sigma}(1+K) \pm \sqrt{\left(\frac{1+K}{2\sigma} - \frac{4L}{\pi G K}\right)^2 \delta G^2 + \left(\frac{4L}{\pi\sigma K} - \frac{G}{2\sigma^2}(1+K)\right)^2 \delta\sigma^2 + \left(\frac{4}{\pi K}\right)^2 \delta L^2}$$

where $K = \sqrt{1 + \frac{16\sigma L}{\pi G}}$. All errors are rounded to one significant figure, though appropriate rounding is only done after the full analysis has been completed. Average error is $\frac{1}{n}\sqrt{\sum \Delta L_i^2}$ where n is the number of voltage levels at each pore or the number of pores considered, as appropriate.

For the sake of comparison, we have included $d_{IV}$, the value of the diameter we measured from the conductance prior to starting the DNA experiment in question (assuming a nominal membrane thickness of 10-nm), as well as $d_{simple}$, which is the value of diameter calculated without access resistance from the conductance blockage according to the equation:

$$\frac{\Delta G}{G} = \frac{d_{DNA}^2}{d_{simple}^2}$$

**Table 2: Averaged values of each nanopore compared to the size extracted from I-V measurements, which assumed a nominal membrane thickness of 10-nm.**

| Name | $L_{average}(nm)$ | $d_{average}(nm)$ | $d_{IV}(nm)$ |
|---|---|---|---|
| **K093** | $8.9 \pm 0.7$ | $6.1 \pm 0.3$ | 6.4 |
| **K097** | $7.2 \pm 0.4$ | $4.6 \pm 0.2$ | 5.3 |
| **K099** | $10.5 \pm 0.9$ | $5.6 \pm 0.3$ | 5.3 |
| **K103** | $10.2 \pm 0.6$ | $6.4 \pm 0.2$ | 5.9 |
| **K104** | $8.8 \pm 0.5$ | $5.8 \pm 0.2$ | 6.1 |
| **K106** | $11 \pm 1$ | $8 \pm 1$ | 7.5 |
| **K123** | $7.6 \pm 0.4$ | $5.6 \pm 0.2$ | 6.2 |
| **K124** | $7.5 \pm 0.4$ | $5.5 \pm 0.2$ | 6 |

Analysis of the DNA data indicates that the membrane thickness at the nanopore location is not particularly thinned by the pore creation process. As importantly, it strongly supports our assumed cylindrical pore geometry as the measured membrane thickness and the nominal membrane thickness generally match within error. This observation is supported by the TEM images shown in Figure S5, revealing a circular opening of the nanopore, consistent with the cylindrical pore model, and not showing signs of membrane thinning around the pore.

**Quantized Conductance Levels –** Further analysis of the translocation data reveals the presence of quantized levels in many blockade events arising from integer multiples of the number of dsDNA strands present in the pores(12, 18). The equidistant separation of the blockade levels of Figure S9 is in agreement with the assumption that the current blockade is proportional to the total cross-sectional area of dsDNA molecules. In addition, we have analyzed each DNA event to measure the average current blockade and their duration. Figure S10 is the resulting scatter plot of the detected translocation events.

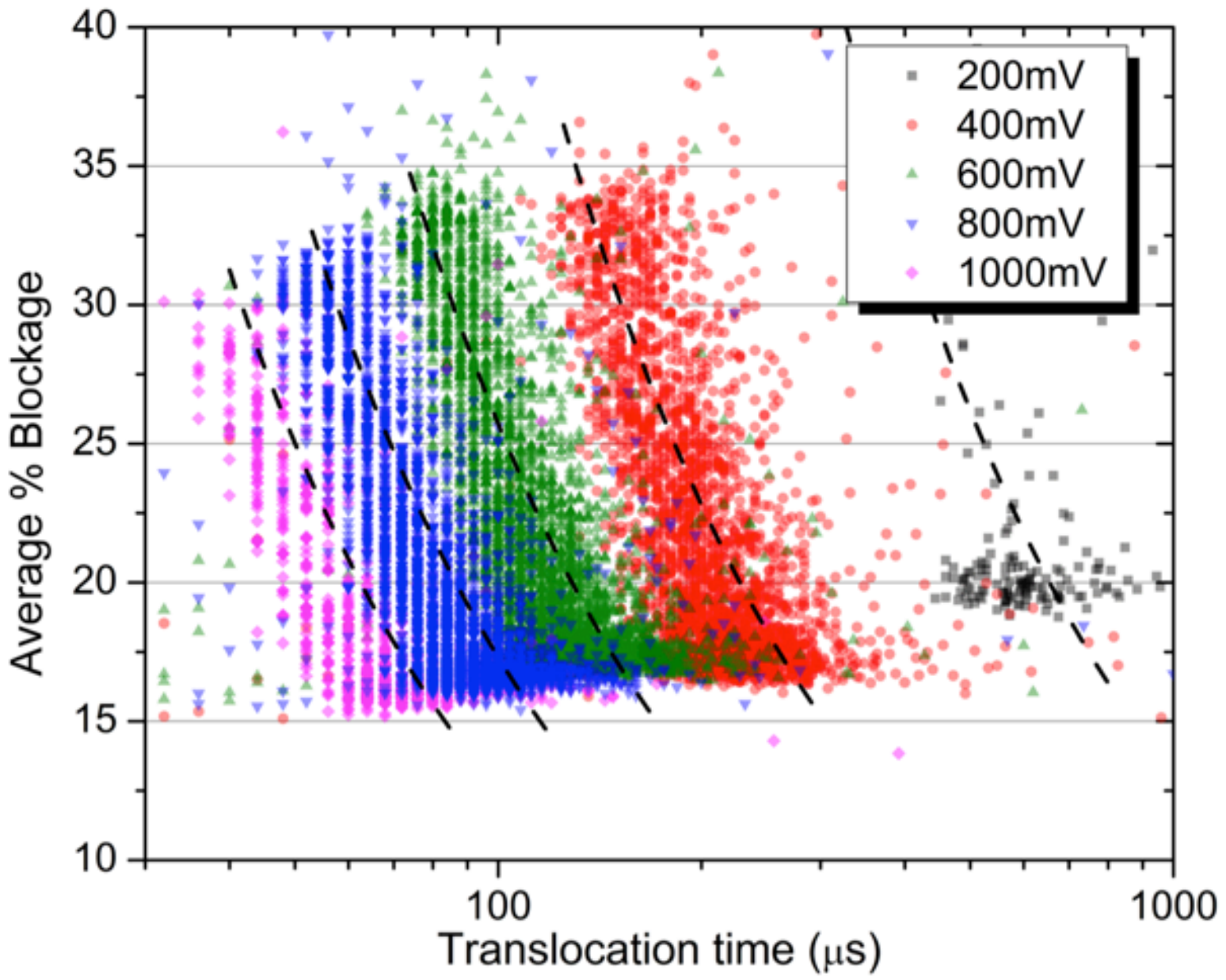

**Figure S10: Scatter plot of normalized ionic current blockade amplitude versus event duration of 5kbp dsDNA translocation events for a 6.5-nm pore in a 10-nm membrane in pH 8 3.6M LiCl. The different behavior at 200mV reflects the fact the pore was ~6-nm and partially clogged after a few hundreds events. A few pulses of moderate electric fields of 0.2V/nm were used unclogged it, which also increased its size to ~6.5-nm. Also note that at 200mV DNA mostly traversed in an unfolded fashion (linear). Data recording at 250kHz sampling rate with a 100kHz 4-pole Bessel filter. The experiments lasted many hours in total. 200mV dataset contains 161 events, 400mV dataset contains 2867 events, 600mV dataset contains 4117 events, 800mV dataset contains 3408 events, and 1000mV dataset contains 747 events. Note that the conductance blockade amplitude at these voltages is slightly reduced, which is also indicative of a larger pore.**

The data agrees reasonably well with the equivalent charge deficit (e.c.d) curve(18, 19), which implies that friction from the wall of a pore is not significant under the conditions tested. Also note that anomalously long DNA translocation events are rare.

The mean translocation time from single-level (unfolded) events at a given trans-membrane potential is extracted by fitting to a Gaussian the distribution to the translocation times. The average DNA velocity is calculated by dividing the length of the DNA by the mean translocation time. Figure S11 shows the mean translocation time and the averaged velocity as a function of applied voltage. The mean translocation time data scale roughly as ~$1/V$ as expected(16). The calculated DNA velocities are in accordance with published results(20), though as with TEM-drilled nanopores we have observed large variability between nanopores.

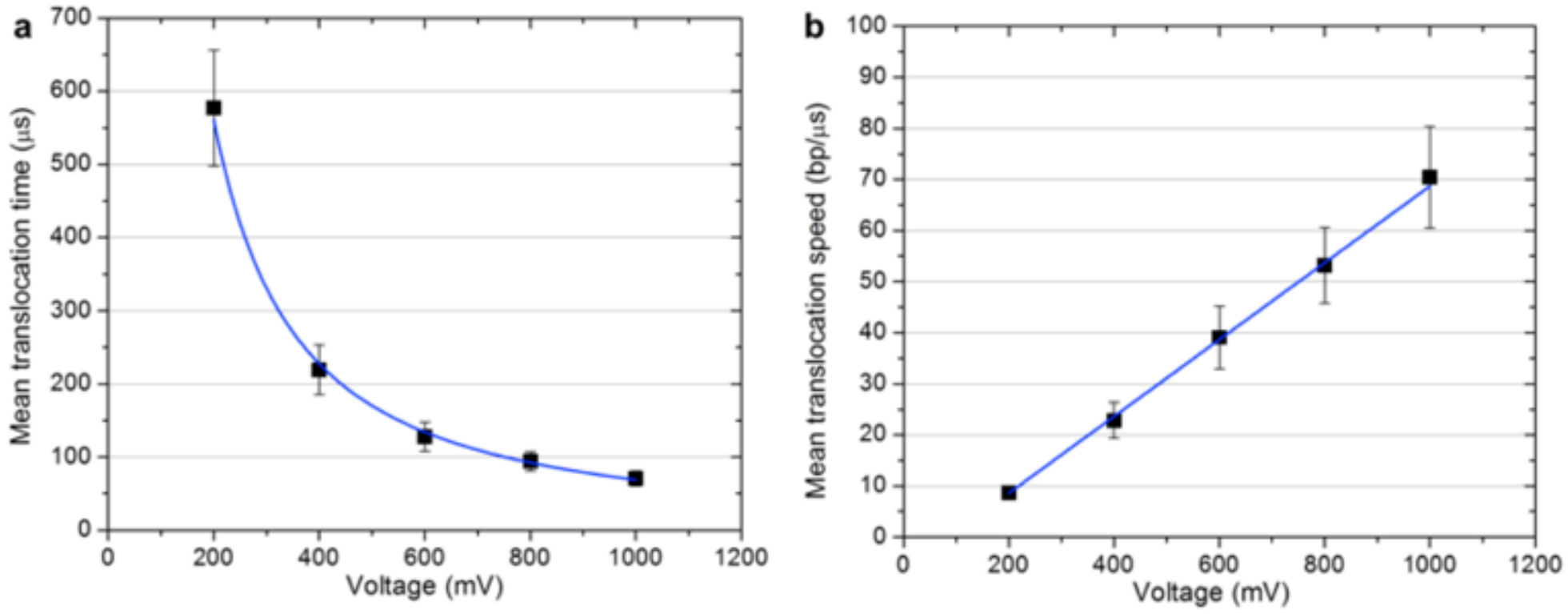

**Figure S11: a) Mean translocation time (dwell time) of single level events (unfolded) versus applied voltage, showing a ~1/V dependence. b) Calculated velocity ($v$ = 5-kbp/dwell time), showing the speed is linear with voltage.**

Finally, the fact that some translocation events characterized 2 and even 3 quantized blockage levels, the latter representing more than 50 % blockage amplitude, strongly supports the conclusion that only a single ~6-nm nanopore is spanning the membrane. If two, or more, nanopores existed on the membrane, considering the size of dsDNA (~2.2nm), 2 and 3 blockade levels could not be achieved, since the area of a single ~6-nm nanopore can be equivalent to two ~4-nm nanopores (assuming cylindrical geometry), and the likelihood of consistently synchronizing translocation events between pores would be low.

Overall, our method fabricates single nanopores, with very high yield, which generate electrical signals from individual translocating DNA molecules that are virtually indistinguishable from TEM-drilled pores. This experimental fact strongly supports the assumption that our fabrication method produces, to first-order, cylindrical channels spanning the membrane, as opposed to long tortuous path or very narrow slits across the membrane. This is further supported by the TEM images obtained of these fabricated pores.

## S9. Strategies to Localize the Pore Creation on the Membrane.

As mentioned in the last paragraph of the manuscript, we suspect that the nanopore creation process to be an intrinsic property of the dielectric membrane, such that a nanopore can form anywhere on its surface. We also provide multiple approaches to easily achieve localization, primarily based on controlling the electric field strength locally on the membrane, since the fabrication time is exponentially related to it (see Figure 3 and Figure S7). Our data suggest that a local change in thickness of 20-nm on the membrane can lead to a $10^4$ change in fabrication time. One can readily envision how such an enormous difference in timescale can be leveraged to localize the pore formation on the membrane.

**RERENCES:**

1. Kimura M, Ohmi T (1996) Conduction mechanism and origin of stress-induced leakage current in thin silicon dioxide films. *J Appl Phys* 80:6360.

2. Lombardo S et al. (2005) Dielectric breakdown mechanisms in gate oxides. *J Appl Phys* 98:121301.

3. Polishchuk I, King T, Hu C (2000) Physical Origin of SILC and Noisy Breakdown in Very Thin Silicon Nitride Gate Dielectric. 4–5.

4. DiMaria DJ, Cartier E (1995) Mechanism for stress-induced leakage currents in thin silicon dioxide films. *J Appl Phys* 78:3883.

5. Tabard-Cossa V, Trivedi D, Wiggin M, Jetha NN, Marziali A (2007) Noise analysis and reduction in solid-state nanopores. *Nanotechnology* 18:305505.

6. Smeets RMM, Keyser UF, Dekker NH, Dekker C (2008) Noise in solid-state nanopores. *Proc Natl Acad Sci U S A* 105:417–21.

7. Beamish E, Kwok H, Tabard-Cossa V, Godin M (2012) Precise control of the size and noise of solid-state nanopores using high electric fields. *Nanotechnology* 23:405301.

8. Smeets RMM, Keyser UF, Wu MY, Dekker NH, Dekker C (2006) Nanobubbles in Solid-State Nanopores. *Phys Rev Lett* 97:1–4.

9. Cui H, Burke P a. (2004) Time-dependent dielectric breakdown of hydrogenated silicon carbon nitride thin films under the influence of copper ions. *Appl Phys Lett* 84:2629.

10. Albella JM, Montero I, Martinez-Duart JM (1987) A THEORY OF AVALANCHE BREAKDOWN OXIDATION DURING. *Electrochim Acta* 32:255–258.

11. Kowalczyk SW, Wells DB, Aksimentiev A, Dekker C (2012) Slowing down DNA translocation through a nanopore in lithium chloride. *Nano Lett* 12:1038–44.

12. Storm a., Chen J, Zandbergen H, Dekker C (2005) Translocation of double-strand DNA through a silicon oxide nanopore. *Phys Rev E* 71:1–10.

13. Bo L, Albertorio F, Hoogerheide DP, Golovchenko JA, Lu B (2011) Origins and consequences of velocity fluctuations during DNA passage through a nanopore. *Biophys J* 101:70–9.

14. Wanunu M et al. (2010) Rapid electronic detection of probe-specific microRNAs using thin nanopore sensors. *Nat Nanotechnol* 5:807–814.

15. Frament CM, Dwyer JR (2012) Conductance-Based Determination of Solid-State Nanopore Size and Shape: An Exploration of Performance Limits. *J Phys Chem C* 116:23315–23321.

16. Fologea D, Uplinger J, Thomas B, McNabb DS, Li J (2005) Slowing DNA translocation in a solid-state nanopore. *Nano Lett* 5:1734–7.

17. Kowalczyk SW, Grosberg AY, Rabin Y, Dekker C (2011) Modeling the conductance and DNA blockade of solid-state nanopores. *Nanotechnology* 22:315101.

18. Li J, Gershow M, Stein D, Brandin E, Golovchenko J a (2003) DNA molecules and configurations in a solid-state nanopore microscope. *Nat Mater* 2:611–5.

19. Fologea D, Brandin E, Uplinger J, Branton D, Li J (2007) DNA conformation and base number simultaneously determined in a nanopore. *Electrophoresis* 28:3186–92.

20. Venkatesan BM, Bashir R (2011) Nanopore sensors for nucleic acid analysis. *Nat Nanotechnol* 6:615–24.